\documentclass[aip,rsi,twocolumn,showpacs,showkeys,preprintnumbers,amsmath,amssymb]{revtex4}
\usepackage{graphicx}
\usepackage{graphics}  
\usepackage{color}  

\usepackage{dcolumn}
\usepackage{bm}

\DeclareGraphicsExtensions{.pdf,.gif,.jpg,.png}

\begin{document}

\newcommand{\oscdemod}{oscillator/demodulator}
\newcommand{\fMUX}{fMUX}
\newcommand{\tMUX}{TDM}
\newcommand{\squid}{SQUID}
\newcommand{\rtHz}{$\sqrt{\mbox{Hz}}$}
\newcommand{\pArtHz}{$\frac{\mathrm{pA}}{\sqrt{\mathrm{Hz}}}$}
\newcommand{\phinot}{\mbox{$\Phi_0$}}
\newcommand{\degree}{\mbox{$^{\circ}$}}
\newcommand{\fortran}{{\tt Fortran~77}}
\newcommand{\CXX}{C++}
\newcommand{\order}{\mbox{${\cal O}$}}
\newcommand{\loopgain}{\mbox{${\cal L}$}}
\newcommand{\const}{\mbox{\sc\small Const}}
\newcommand{\approxlt}{ \stackrel{<}{\sim} }
\newcommand{\approxgt}{ \stackrel{>}{\sim} }
\newcommand{\tauweb}{\tau_\mathrm{web}} 
\newcommand{\tauTES}{\tau_\mathrm{TES}^\mathrm{eff}}      
\newcommand{\tauLCR}{\tau_\mathrm{LCR}}   
\newcommand{\Rbolo}{R_\mathrm{bolo}}
\newcommand{\Rnormal}{R_\mathrm{normal}}
\newcommand{\Prad}{P_\mathrm{rad}}
\newcommand{\Pelec}{P_\mathrm{elec}}

\newcommand{\valGammaNE}{0.5}

\newcommand{\mycomment}[1]{$\dagger$ \marginpar{\it\tiny $\dagger$ #1}} 
\newcommand{\mynotes}[1]{\newline {\footnotesize {\tt #1 } } }
\newcommand{\mynewtext}[1]{ #1 }
\newcommand{\revisedtext}[1]{ #1 }
\newcommand{\partno}[2]{{ #1 (#2)}} 

\newcommand\arcdeg{\mbox{$^\circ$}}%
\newcommand\arcmin{\mbox{$^\prime$}}%
\newcommand\arcsec{\mbox{$^{\prime\prime}$}}%

\newcommand\onehalf{\mbox{$\frac{1}{2}$}}%
\newcommand\onethird{\mbox{$\frac{1}{3}$}}%
\newcommand\twothirds{\mbox{$\frac{2}{3}$}}%
\newcommand\onequarter{\mbox{$\frac{1}{4}$}}%
\newcommand\threequarters{\mbox{$\frac{3}{4}$}}%


\title{Frequency multiplexed superconducting quantum interference device readout \\ 
of large bolometer arrays for cosmic microwave background measurements}

\newcommand{\mcgill}{1}
\newcommand{\ucb}{2}
\newcommand{\caltech}{3}
\newcommand{\chicagoonly}{4}
\newcommand{\colorado}{5}
\newcommand{\kavli}{6}
\newcommand{\fermi}{7}
\newcommand{\chicagophys}{8}
\newcommand{\chicagoastro}{{9}}
\newcommand{\nist}{{10}}
\newcommand{\lbnlmaterials}{{11}}
\newcommand{\columbia}{{12}}
\newcommand{\lbnleng}{{13}}
\newcommand{\dwave}{{14}}
\newcommand{\lbnlphys}{{15}}
\newcommand{\michigan}{{16}}
\newcommand{\casewestern}{{17}}
\newcommand{\minnesota}{{18}}
\newcommand{\artinstitute}{{19}}
\newcommand{\hsca}{{20}}

\author{M.A.\ Dobbs$^\mcgill$, 
M.\ Lueker$^{\ucb,\caltech}$,
K.A.\ Aird$^\chicagoonly$, 
A.N.\ Bender$^\colorado$, 
B.A.\ Benson$^{\kavli,\fermi}$,
L.E.\ Bleem$^{\kavli,\chicagophys}$, 
J.E.\ Carlstrom$^{\kavli,\chicagophys,\fermi,\chicagoastro}$, 
C.L.\ Chang$^{\kavli,\fermi}$, 
H.-M.\ Cho$^\nist$,
 J.\ Clarke$^{\ucb,\lbnlmaterials}$, 
T.M.\ Crawford$^{\kavli,\chicagoastro}$, 
A.T.\ Crites$^{\kavli,\chicagoastro}$, 
D.I.\ Flanigan$^\ucb$,
T.\ de Haan$^\mcgill$, 
E.M.\ George$^{\ucb}$, 
N.W.\ Halverson$^\colorado$, 
W.L.\ Holzapfel$^{\ucb}$, 
J.D.\ Hrubes$^\chicagoonly$,
B.R.\ Johnson$^{\ucb,\columbia}$,
J.\ Joseph$^\lbnleng$, 
R.\ Keisler$^{\kavli,\chicagophys}$, 
J.\ Kennedy$^\mcgill$, 
Z.\ Kermish$^{\ucb}$, 
T.M.\ Lanting$^{\mcgill,\dwave}$, 
A.T.\ Lee$^{\ucb,\lbnlphys}$, 
E.M.\ Leitch$^{\kavli,\chicagoastro}$, 
D.\ Luong-Van$^\chicagoonly$, 
J.J.\ McMahon$^\michigan$, 
J.\ Mehl$^{\kavli,\chicagoastro}$, 
S.S.\ Meyer$^{\kavli,\chicagophys,\fermi,\chicagoastro}$,
T.E.\ Montroy$^\casewestern$, 
S.\ Padin$^{\caltech,\kavli,\fermi}$, 
T.\ Plagge$^{\kavli,\chicagoastro}$, 
C.\ Pryke$^\minnesota$, 
P.L.\ Richards$^{\ucb}$, 
J.E.\ Ruhl$^\casewestern$,
K.K.\ Schaffer$^{\artinstitute,\kavli,\fermi}$,
D.\ Schwan$^{\ucb}$, 
E.\ Shirokoff$^{\ucb,\caltech}$, 
H.G.\ Spieler$^\lbnlphys$,
Z.\ Staniszewski$^{\casewestern,\caltech}$, 
A.A.\ Stark$^\hsca$, 
K.\ Vanderlinde$^\mcgill$, 
J.D.\ Vieira$^\caltech$, 
C.\ Vu$^\lbnleng$, 
B.\ Westbrook$^{\ucb}$, 
R.\ Williamson$^{\kavli,\chicagoastro}$ }



\affiliation{
$^\mcgill$Physics Department, McGill University, Montreal, Quebec H3A~2T8, Canada \\
$^\ucb$Physics Department, University of California, Berkeley, California 94720, USA \\
$^\caltech$Division of Physics, Math and Astronomy, California
Institute of Technology, 1200 E. California Blvd., Pasadena, CA 91125 \\
$^\chicagoonly$University of Chicago, 5640 South Ellis Avenue,
Chicago, IL 60637 \\
$^\colorado$Department of Astrophysical and Planetary Sciences and Department
of Physics, University of Colorado, Boulder, CO 80309 \\
$^\kavli$Kavli Institute for Cosmological Physics, University of Chicago, 5640 South Ellis Avenue, Chicago, IL 60637 \\
$^\fermi$Enrico Fermi Institute, University of Chicago, 5640 South
Ellis Avenue, Chicago, IL 60637 \\
$^\chicagophys$Department of Physics, University of Chicago, 5640 South Ellis
Avenue, Chicago, IL 60637 \\
$^\chicagoastro$Department of Astronomy and Astrophysics, University of
Chicago, 5640 South Ellis Avenue, Chicago, IL 60637 \\
$^\nist$National Institute of Standards and Technology, Boulder,
Colorado 80305, USA \\
$^\lbnlmaterials$Materials Science Division, Lawrence Berkeley National Lab, Berkeley,
California 94720, USA \\
$^\columbia$Department of Physics, Columbia University, 538 West 120th Street, New
York, NY 10027 \\
$^\lbnleng$Engineering Division, Lawrence Berkeley National Lab, Berkeley,
California 94720, USA \\
$^\dwave$ now at D-Wave Systems, Burnaby, British Columbia V5C~6G9,
Canada \\
$^\lbnlphys$Physics Division, Lawrence Berkeley National Lab, Berkeley,
California 94720, USA \\
$^\casewestern$Physics Department and CERCA, Case Western Reserve University,
10900 Euclid Ave., Cleveland, OH 44106 \\
$^\michigan$Department of Physics, University of Michigan, 450 Church
Street, Ann Arbor, MI, 48109 \\
$^\minnesota$University of Minnesota/Twin Cities, School of Physics and
Astronomy, Minneapolis, MN, 55455 \\
$^\artinstitute$Liberal Arts Department, School of the Art Institute of Chicago, 112 S Michigan Ave, Chicago, IL 60603 \\
$^\hsca$Harvard-Smithsonian Center for Astrophysics, 60 Garden
Street, Cambridge, MA 02138
}
\date{\today}

\begin{abstract}  

  A technological milestone for experiments employing Transition Edge
  Sensor (TES) bolometers operating at sub-kelvin temperature is the
  deployment of detector arrays with
  100s--1000s of bolometers. 
  One key technology for such arrays is readout
  multiplexing: the ability to read out many sensors simultaneously on
  the same set of wires. This paper describes a frequency-domain
  multiplexed readout system which has been developed for and deployed on
  the APEX-SZ and South Pole Telescope millimeter wavelength
  receivers. In this system, the detector array is divided into
  modules of seven detectors, and each bolometer within the module is
  biased with a unique $\sim$MHz sinusoidal carrier such that the
  individual bolometer signals are well separated in frequency space. The
  currents from all bolometers in a module are summed together and
  pre-amplified with Superconducting Quantum Interference Devices
  (SQUIDs) operating at 4\,K. Room-temperature electronics demodulate the carriers
  to recover the bolometer signals, which are digitized separately and
  stored to disk. This readout system contributes little noise
  relative to the detectors themselves, is remarkably insensitive to
  unwanted microphonic excitations, and provides a technology pathway
  to multiplexing larger numbers of sensors.
\end{abstract}


\maketitle

\section{Introduction}

A new generation of mm-wavelength receivers instrumented with
hundreds or thousands of transition edge sensor (TES) bolometric
detectors, e.g.,
\cite{2011ApJS..194...41S,apexSZ_instrument,2010ApJ...711.1141T,2009arXiv0907.4445C,2008AIPC.1040...66L},
%
%
is allowing for unprecedented precision in measurements of the Cosmic
Microwave Background (CMB). Large TES arrays are expected to bring forth
similar advances in other observational bands such as the sub-mm
(e.g., \cite{2006SPIE.6275E..45H}), X-ray~(e.g.,
\cite{2008JLTP..151..363S}), and gamma-ray~(e.g.,
\cite{2007ApPhL..90s3508D,2004ApPhL..85.4762D,dreyer_llnl2007})
as well as phonon detection in dark matter
direct detection experiments (e.g., \cite{Wang:1990qk}).

TES detectors in the millimeter to far-IR range are approaching the
photon statistics limit, meaning their noise performance is dominated
by fluctuations in the arrival rate of photons at the detector, 
and not by the intrinsic noise performance of the detectors themselves. 
Consequently, higher mapping speed can best be realized by
increasing the number of detectors, rather than by improving the
detectors themselves. A key challenge in building focal planes with
large sensor counts has been multiplexing the signals between the cold
sub-kelvin operating temperature of the detectors and room temperature
electronics. Multiplexing minimizes the heat-load on the detector cold
stage and reduces the complexity of the cold wiring.

Two complementary multiplexing strategies are commonly used: (1) time
domain multiplexing (\tMUX)~\cite{chervenak99,2000NIMPA.444..107C,
  2002AIPC..605..301I,2008JLTP..151..908B}, where the detectors are
read out sequentially one at a time with a revisit rate that is faster
than the detector time constant, and (2) frequency-domain multiplexing
(\fMUX) where each detector is biased at a unique location in
frequency space and read out continuously. While the \tMUX\ system was
developed earlier, both technologies are now quite mature. 
\mynewtext{
An eight-channel \tMUX\ proof-of-concept system was first to make sky observations~\cite{2003SPIE.4855..100S}.
}
A new TES readout
scheme that multiplexes signals from GHz frequency RF \squid
s~\cite{2004ApPhL..85.2107I,2006NIMPA.559..802I} is in the early
stages of development, as is a new code-domain multiplexing
readout~\cite{2010SuScT..23c4004I}. 

In this paper, we describe the \fMUX\ readout system
\cite{yoon01,2002AIPC..605..305Y,Spieler02,2003SPIE.4855..172L,2004NIMPA.520..548L,2005ApPhL..86k2511L,2006NIMPA.559..793L,2006PhDT........32L}
as it has been deployed for mm-wavelength observations with the
APEX-SZ~\cite{apexSZ_instrument,2006NewAR..50..960D,2003NewAR..47..933S}
and South Pole Telescope (SPT)
SZ~\cite{2009arXiv0907.4445C,2004astro.ph.11122R} receivers. 
\mynewtext{
These
experiments are designed for high angular resolution observations of
the CMB.  For example, the unique combination of sensitivity and
resolution of 
these instruments makes it possible to study the structure and
evolution of galaxy clusters using the Sunyaev-Zel'dovich (SZ)
effect~\cite{1972CoASP...4..173S}, a small distortion imprinted on the
CMB when primordial photons inverse-Compton scatter off hot electrons
in the intra-cluster medium.
}  The resolved SZ image of the Bullet
Cluster (1E 0657-56) obtained with APEX-SZ~\cite{2008arXiv0807.4208H}
was the first published scientific result with a large array of
multiplexed TES detectors. SPT achieved the first SZ detections of
previously unknown galaxy clusters~\cite{2008arXiv0810.1578S} with its
array of multiplexed TES detectors. 
\mynewtext{
A wealth of new measurements 
have been presented by both SPT~\cite{
2011arXiv1111.0932R,
2011ApJ...743...28K,2011ApJ...735L..36S,2011ApJ...738..139W,
2011ApJ...736...61S,2010ApJ...721...90B,2011ApJ...738...48A,
2010ApJ...722.1180V,2010ApJ...723.1736H,2010ApJ...719.1045L,
2010ApJ...718..632H,2010ApJ...719..763V,2010ApJ...716.1118P} 
and APEX-SZ~\cite{2010A&A...519A..29B,2009ApJ...701.1958R,2009A&A...506..623N}.
}

Other implementations of frequency multiplexed TES readout systems are
described in Refs.~\cite{2009AIPC.1185..245V,2008ITNS...55...21D}.


This paper is outlined as follows. In \S\ref{s_general_overview}, a
brief overview of the system, its advantages, and the performance
requirements is presented. A detailed description of the sub-kelvin
components of the system is provided in \S\ref{s_cold_mux}. This
description is accompanied by a summary of the design process for
specifying the basic multiplexer parameters, such as channel spacing.
\S\ref{s_squid_and_electronics} details the \squid\ pre-amplifier and
back-end electronics. The performance for the system operating in the
SPT and APEX-SZ experiments is described in \S\ref{s_performance}, and
conclusions are presented in \S\ref{s_conclusions}.

\section{General Overview} \label{s_general_overview}

The \fMUX\ readout system operates in concert with an array of TES
detectors. Each bolometric detector has a metal absorber and
TES, which are coupled by a weak thermal link to a $\sim 0.25$\,K
thermal bath. A constant-amplitude sinusoidal voltage bias is applied
to the TES, which is a superconducting film with a transition
temperature of about 0.5\,K. The combination of radiation power
absorbed from the sky and electrical bias power raises the temperature
of the TES mid-way into its superconducting transition.
\revisedtext{An increase in sky power incident on the absorber alters the TES
   resistance and (through negative electro-thermal feedback, see
   \S\ref{s_TES_bolometers}), produces a decrease in the electrical power
   being dissipated in the sensor.} This, in turn, produces a change
in the current through the sensor, since the applied voltage is held
constant.
TES thermometers are chosen because their steep transition allows
for linearity, their monolithic wafer-scale fabrication yields large arrays,
and because their low impedance makes them
compatible with multiplexing. The properties and parameters of the
bolometer arrays dictate the design requirements for the readout
system.

\subsection{The APEX-SZ and SPT Experiments}

The SPT focal plane~\cite{BensonSptReceiver, 2008_ErikSPTDetectors} is
divided into three separate bands centered at approximately 95, 150,
and 220\,GHz, with bandwidths of about 40\,GHz. The APEX-SZ detectors
operate at 150\,GHz with $\sim$36\,GHz bandwidth. Coherent
amplification devices such as High Electron Mobility Transistors
(HEMTs) are ultimately limited by quantum noise, and incoherent
detection with bolometers can offer higher sensitivity at these
frequencies.

Both experiments use a 
closed-cycle refrigerator manufactured by \partno{Chase Research
  Cryogenics}{140 Manchester Road, Sheffield, UK S10 5DL}, with two
$^{3}$He sorption stages and one $^{4}$He sorption stage, to cool the
bolometers and sub-kelvin multiplexer components to their
$\sim$0.25\,K bath temperature. The refrigerator is thermally coupled to the
$\sim$4\,K stage of a \partno{Cryomech}{113 Falso Drive, Syracuse, USA
  13211} model PT\,410 mechanical pulse tube cooler. With this
combination, the system is cooled without the use of any expendable
cryogens, greatly simplifying the operation of these instruments in
the remote environments of the Chilean Atacama plateau and the
geographic South Pole for APEX-SZ and SPT respectively.

APEX-SZ and SPT have 0.4$^\circ$ and 1$^\circ$ fields of views (FoVs)
respectively, and both experiments have full width half maximum (FWHM)
beam sizes of about 1~arcminute. There are no ``chopper mirrors'' in
the optical systems to modulate the beams across the sky; instead the
telescopes are scanned at typical speeds of 0.25-0.5$^\circ$/second.
Since we are interested in observing astronomical structures (such as
primary CMB fluctuations) at scales occupying a large fraction of the
FoV, these experimental parameters and scan strategy dictate that
information will appear in the detector timestreams at frequencies as
low as $\approxlt 1$~Hz.

\subsection{Readout System Concept}
\label{s_system_concept}

\begin{figure}
\includegraphics[width=0.5\textwidth,clip=]{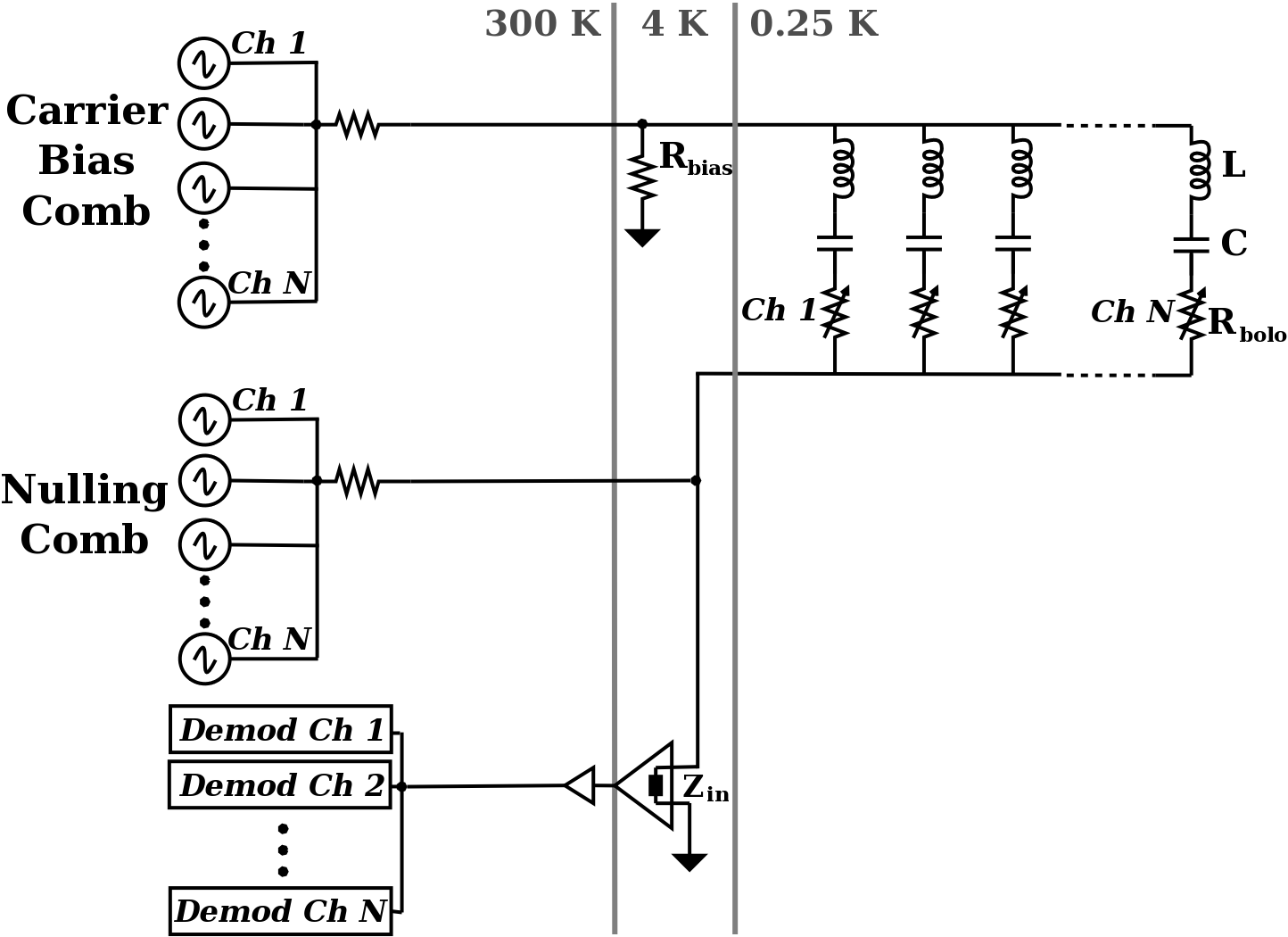}
\caption{ \label{f_fmux_cartoon}
Layout of the frequency-domain multiplexed readout for 
a module of $N$ TES detectors.}
\end{figure}

\begin{figure*} 
 \includegraphics[width=\textwidth,clip=]{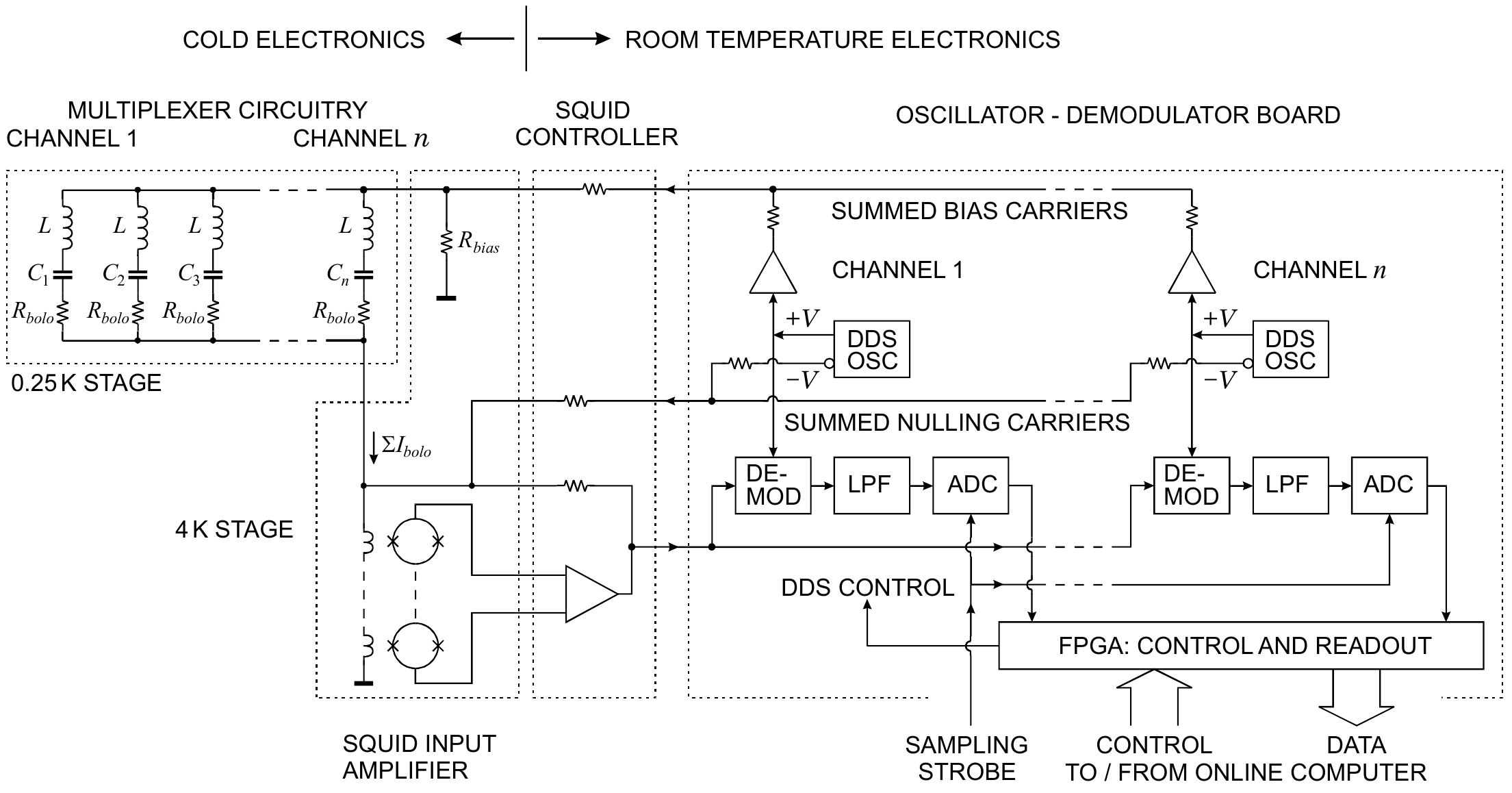}
 \caption{ \label{f_fmux_implementation} Block diagram showing the
   basic topology of the \fMUX\ readout system implementation, with
   an emphasis on the room temperature electronics. Refer to
   \S\ref{s_squid_and_electronics} for a description.}
\end{figure*}

The detector readout is segregated into identical \fMUX\ modules, each
biasing and reading out $N$ TES detectors on a single pair of wires.
The basic layout of one multiplexer module of the \fMUX\ system is
shown in Figure~\ref{f_fmux_cartoon}.
Figure~\ref{f_fmux_implementation} shows the system electronics
(\S~\ref{s_squid_and_electronics}) in greater detail.  To set its
specific bias frequency, each TES bolometer ($\Rbolo$) is connected
through a series-resonant circuit formed by an inductor ($L$) and
capacitor ($C$) on the sub-kelvin stage. The inductance is the same
for all channels, defining the same $\Rbolo/(2\pi L)$ electrical
bandwidth for each detector.  \mynewtext{ Thus, changing the
  capacitance alone sets the $LC\Rbolo$ channel resonant frequency. }
A bank of $N$ fixed amplitude sine-wave generators operating at room
temperature provides a `comb' of detector voltage bias carriers, each
operating at a frequency tuned to match the resonance of the
individual $LC\Rbolo$ channels. Sky signals modulate the TES
resistance, amplitude modulating the current of the carrier. This
encodes the sky-signal in symmetrical sidebands above and below the
carrier.
\revisedtext{
  Since the signal currents from the individual TES detectors are at
  different frequencies, they can all be summed together and transmitted to a
  single low-impedance input amplifier}
operating at 4\,K. We
employ a series array \squid\ operating in a flux-locked loop (FLL)
for this transimpedance amplification stage. 
\revisedtext{
  To reduce the \squid\ dynamic range requirements, an inverted
  version of the carrier comb, referred to as the `nulling comb', is
  injected at the \squid\ input to cancel the carriers (see
  \S\ref{s_squid_and_electronics}).}
The comb of amplitude
modulated carriers is transmitted from the \squid\ output to a bank of
room temperature demodulators. There is separate analog demodulator
for each TES that mixes the detector signal down to baseband. A
low-pass anti-aliasing filter is applied to the resultant timestream
before it is digitized. The signal amplitude is proportional to the
photon power deposited on the TES photon absorber. The APEX-SZ and SPT
systems are configured with $N=7$ detectors per readout module.


The configuration shown in Figure~\ref{f_fmux_cartoon} uses current
summing at the low impedance input of the transimpedance amplifier
using a \squid\ input stage. The \fMUX\ can also be implemented by
introducing a transformer in series with each bolometer, and summing
the voltages from the transformer
secondaries~\cite{2002AIPC..605..305Y}, but since a TES
biased by a constant voltage yields changes in signal current, the
current summing configuration is more direct.  The design
requirements, challenges, and implications are similar for both
configurations. For simplicity, we focus only on the current summing
strategy.

The \fMUX\ readout architecture has the following key features:
\\ \noindent
\mynewtext{
(1) Simplicity of cryogenic components: the sub-kelvin components
consist of one capacitor and one superconducting inductor per bolometer, 
followed by a single \squid\
pre-amplifier per readout module at 4K. Since the \squid s are located far from the
detectors, they can be magnetically shielded in a straightforward
manner. The only custom sub-kelvin components are the inductors,
which can be manufactured via a simple, low cost process with high yield. 
}
\\ \noindent
(2) Low wire count per detector and independent modules. 
The \fMUX\
requires just two wires per readout module for both bias and readout.
Each module is completely independent so that wiring failures or
other defects in one module do not affect others.
\\ \noindent (3) Signals are modulated at high (typically MHz)
frequencies, well above microphonic pickup frequencies 
and amplifier low-frequency noise.
\revisedtext{ 
  The inductor-capacitor filters in series with the detectors suppress 
  currents at frequencies outside the narrow LCRbolo bandwidth of the 
  individual bolometer legs. 
  This reduces crosstalk, eliminates the dominant source of
  microphonics usually present in bolometer systems, and results in a
  high rejection of electromagnetic interference that comes from
  numerous sources such as scan-correlated changes in magnetic field.}
\\ \noindent (4) The readout system dissipates zero power on
the detector cold stage as there are no resistive or active
elements. 
\\ \noindent (5) Channels are read out continuously with no
switching transients that limit the detector bandwidth. This makes the
system easily applicable to higher bandwidth application such as
photon counting X-ray microcalorimeters. 
\\ \noindent (6) The voltage bias can be set to optimize
the sensitivity of each individual detector. This allows the readout
system to compensate for detector fabrication non-uniformity.

\mynewtext{
Practical challenges for implementing the \fMUX\ system include:
achieving the large dynamic range needed to accommodate the bias
carriers; meeting the strict stability requirements on the bias
generators needed to avoid introducing low frequency noise; managing
non-idealities in cryogenic wiring that can introduce stray inductances
that spoil voltage bias or create cross-talk; avoiding detector
oscillations which can affect other detectors in the same readout
module through \squid\ interactions; and, since the \squid\ FLL
includes a room temperature amplifier, achieving loop stability in the
present implementation of this transimpedance amplifier is contingent
on short wire lengths between 4\,K and room temperature.
}

\subsection{Design Requirements} \label{s_design_requirements}
\newcommand{\smallheading}[1]{\\ \noindent {\it #1}:}

The TES detector properties, our telescope scan strategy, and our science goals 
define the following requirements for the \fMUX\ readout system:
\smallheading{Noise}
 The noise contribution from the readout system should be 
 small compared to photon and detector noise which is
 typically about 50~aW/\rtHz\ for ground-based mm-wavelength observations. 
 The readout system receives noise
 contributions from its \squid\ and room-temperature electronics amplifier
 stages, Johnson noise from the TES resistance, and Johnson noise
 leakage through the $LC\Rbolo$ filter from neighboring detector channels.
\smallheading{Detector Stability}
 The analog circuit that includes the TES must be stable across the full 
 range of observation conditions.
\smallheading{Cross-talk}
\mynewtext{
 Typical detector-to-detector cross-talk for our systems due to beam side-lobes
 and radiation leakage in the bolometer integration cavity
 is $\sim 1$\%.} The readout system electrical cross-talk should be
 small in comparison.
\smallheading{Dynamic Range}
 The readout system amplification needs to have sufficient dynamic
 range to accommodate the large bias carriers
 without introducing significant intermodulation
 distortion. The dynamic range requirement is substantially reduced
 for the \fMUX\ system by canceling the carriers at the \squid\ input 
 with nulling signals that are inverted versions of the raw bias carriers,
described in \S\ref{s_squid_and_electronics}.
\smallheading{Bandwidth and Post-Detection Sampling Rate}
 The telescope scan strategy defines a low frequency stability
 requirement of better than $\sim$1~Hz which places stringent
 stability requirements on the sideband noise of the bias carriers.
 \revisedtext{ The fastest expected thermalization time constant of
   the detector absorber, 
   $\tauweb\simeq$5\,ms (often referred to as the ``optical time constant''),  
   defines 30~Hz as the
   upper end of the signal band. Note that much higher bandwidths are
 of interest for characterizing bolometers, since the response time of the TES
 thermistor $\tauTES$ can be much faster than the bolometer
 optical time constant.} This motivates a sampling rate in the range
 of 100-1000~Hz. 
 \revisedtext{Refer to Table~\ref{t_timeconstants} for definitions of
    the time constants used in this paper and Ref.~\cite{2009ITAS...19..496L} for a
   detailed thermal description of the detectors, including time
   response and diagrams.}
\smallheading{Programmability} The system needs to operate 
 autonomously under computer control, tuning and setting up each
 \squid\ and detector. This tuning needs to be dynamic, so that
 changes in operating conditions (such as small changes in cryogenic
 temperatures or large changes in atmospheric loading) do not
 adversely affect the system yield or operation.
\smallheading{Diagnostics} The system must provide sufficient 
 diagnostic information so as to characterize detector performance,
 recognize problems such as failed tunings, and provide feedback to
 the telescope operator.


\revisedtext{
\begin{table}
\begin{tabular}{l p{0.45\textwidth} } \hline
 $\tauweb$ & thermalization time constant of the detector radiation absorber  \\
 $\tauTES$  & effective TES response time for thermal signals \\
 $\tauLCR$  & electrical time constant of the notch filter formed by the bolometer
                       resistance and multiplexer inductor/capacitor \\ \hline
\end{tabular}
\caption{\label{t_timeconstants}
Definitions for the time constants utilized in this paper.
}
\end{table}
}

\section{Sub-Kelvin Multiplexer} \label{s_cold_mux}

One of the advantages for the \fMUX\ system is the simplicity of the
sub-kelvin readout components, consisting only of capacitors,
inductors, and wiring.
However, as discussed in this section, successful implementation of an
\fMUX\ system requires careful management of stray reactances to
maintain good detector stability and optimal system noise. In addition
to describing the specific implementations of the sub-kelvin
multiplexer components for APEX-SZ and SPT in this section, 
after first describing the TES bolometers,
we outline the design process that leads to the choice of parameters.

\subsection{TES Bolometer Arrays} \label{s_TES_bolometers}

\begin{figure}
\includegraphics[height=0.45\textwidth,clip=,angle=90]{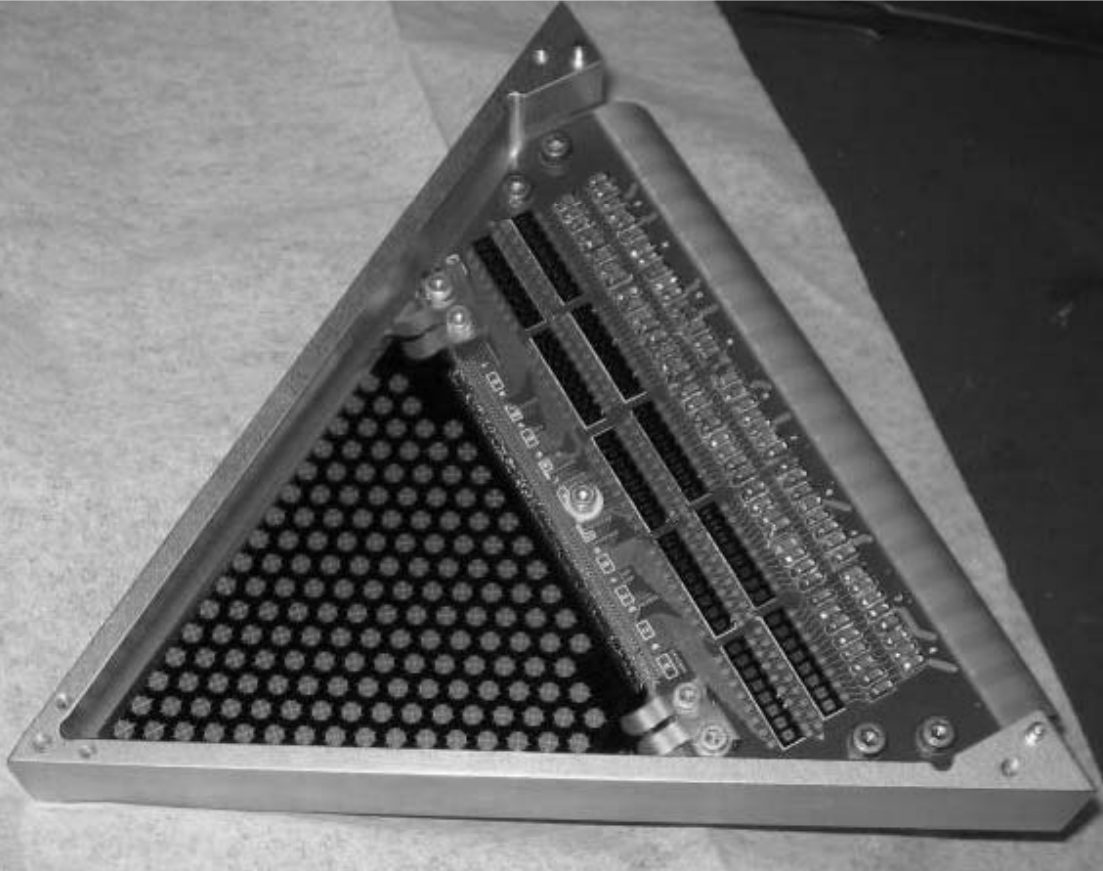}
\caption{ \label{f_bolometer_photo} Photograph showing one wedge of
161 SPT TES bolometers. The bolometer wedge is wire-bonded to a
circuit board on which the multiplexer inductors and capacitors are
mounted. An integrated flexible microstrip (visible at the right of
the photo) brings the leads to connectors mounted below the
circuit board. The band-defining feed-horns and metal mesh filters
have been removed for this photo.
}
\end{figure}

The APEX-SZ and SPT TES bolometer arrays
(Figure~\ref{f_bolometer_photo},) were fabricated at UC
Berkeley~\cite{2008JLTP..151..697M,2008_ErikSPTDetectors,2011ShirokoffPhD}. The
detector arrays consist of six triangular shaped wedges, which together
form a hexagonal array of 330 and 966 bolometers respectively for
APEX-SZ and SPT. Short sections of circular wave-guide with 
conical feed-horns define the lower edge of the
detector band and couple the detectors to free space. Metal mesh
filters~\cite{2006SPIE.6275E..25T,2006SPIE.6275E..26A} located above
the feed-horns define the upper band edge. The bolometers are electrically
connected to the cold multiplexer components with aluminum wire bonds
on the periphery of the bolometer array.

The bolometers each consist of a 3\,mm diameter gold absorber suspended
above the 0.26\,K cold stage bath through a thermal conductance in the
range of $\bar{G}$=100-200\,pW/K, where the bar indicates $G$ is
averaged across the temperature difference between the TES transition
temperature and its thermal bath. The bolometer operates with total
power $P_\mathrm{total} = \bar{G} (T_\mathrm{bolo}-T_\mathrm{bath})$
flowing to the bath. 
Optimizing $G$ is a trade-off between (1) lower noise, which pushes
$G$ to lower values, and (2) high dynamic range for observations at a
variety of optical loading conditions, which pushes $G$ to higher
values. For 
APEX-SZ and SPT, the target $G$ has been chosen so that the detector
saturation power is slightly larger than twice the typical incident
radiation power, $P_\mathrm{total} \gtrsim 2\,P_\mathrm{rad}$, which
maintains high 
electro-thermal loop-gain (discussed below) under all observing
conditions.

The bolometer absorber has a spiderweb geometry, which provides a high
optical efficiency while minimizing the cross section to cosmic rays.
An Al/Ti bilayer TES is located at the center of each spiderweb and
tuned to a superconducting transition temperature $T_\mathrm{bolo} \sim
0.55$\,K and normal resistance $\Rnormal \sim 1.2\,\Omega$. 
\revisedtext{
  The thermalization time of the spiderweb $\tau_{web}$, typically 10-15
  ms (16-11 Hz) for our detectors, sets the device's response time for optical signals.}
\revisedtext{ The TES has a response time $\tauTES$ to thermal signals
  that is sped up by electro-thermal feedback and is typically more
  than an order of magnitude faster than $\tauweb$. }
Since a fast $\tauTES$ places additional constraints on 
the readout system (see \S\ref{s_stability}), 
\revisedtext{the TES time constant is slowed down by attaching a gold
  feature with considerable heat capacity} 
\cite{2009ITAS...19..496L}. Typical optical efficiencies
(for the conversion of power from astronomical radiation to electrical
power) are $\sim$80\% for light incident on the detectors, and
about 20-40\% end-to-end from the telescope's primary dish. References
\cite{apexSZ_instrument} and \cite{BensonSptReceiver}
provide discussions of the APEX-SZ and SPT optical efficiences respectively. 


\subsubsection{Detector Bias and Electro-Thermal Feedback}

\mynewtext{
The TES are biased with a fixed amplitude sinusoidal voltage
at frequencies much higher (faster) than the TES can respond, so it sees the integrated bias power. }
The steep $\Delta \Rbolo/\Delta T$ TES transition and
voltage bias provide the conditions for strong
negative electro-thermal feedback (ETF)~\cite{irwin:1998}. When the
radiation power $P_\mathrm{rad}$ incident on the bolometer increases, the
bolometer temperature $T_\mathrm{bolo}$ increases, driving up its resistance
$\Rbolo$. This, in turn, brings down the electrical bias power
$P^{\omega^i}_\mathrm{elec} = V_\mathrm{bias}^2 / \Rbolo$
provided by the sinusoid at frequency $\omega^i$, completing the
negative ETF loop. 

\revisedtext{ETF improves linearity, extends the dynamic range, and
  speeds up the effective TES response time for thermal signals.}
A measure of the ETF strength is the loop-gain,
\begin{equation} \label{e_etf}
\loopgain_i = \frac{\alpha P^{\omega^i}_\mathrm{elec}}{G T_\mathrm{bolo}}, 
\end{equation}
where $\alpha = d \log (\Rbolo) / d \log (T_\mathrm{bolo})$ is the logarithmic
derivative of the bolometer resistance $R_\mathrm{bolo}$ with temperature
$T_\mathrm{bolo}$,
\revisedtext{
$G=d P_\mathrm{total}/d T_\mathrm{bolo}$}
 is the instantaneous thermal conductance
between the bolometer absorber and bath, and $T_\mathrm{bolo}$ is the TES
transition temperature. When the loop-gain is high,
the total power on the bolometer 
$P_\mathrm{total} = P_\mathrm{rad} + P^{\omega^i}_\mathrm{elec}$ will be
approximately constant and balanced by the rate of heat transmitted to
the bath,
\revisedtext{ $P_\mathrm{total} = \bar{G} (T_\mathrm{bolo}-T_\mathrm{bath})
                         =\int_{T_\mathrm{bath}}^{T_\mathrm{bolo}} G dT$ .}



One simple strategy for providing voltage bias is to place the TES in
a loop with a small bias resistor, $R_\mathrm{bias}<<\Rbolo$ shown in
Figure~\ref{f_fmux_cartoon}, across which the voltage bias is provided
by an external current source. 
This loop consists of the bias resistor, one
bolometer, the inductor and capacitor that define the loop resonant
frequency, and the transimpedance amplifier input.
Ideally, all other components in the
loop would have negligible impedance in comparison with the bolometer.
\revisedtext{
  This includes the transimpedance amplifier that measures the TES
  current, wiring, and any other components such as the inductors and
  capacitors that define the resonant frequency. }
The wiring that connects the bolometers on the sub-kelvin stage to the
low impedance ammeter (in this implementation, a \squid\ array located
on a  4\,K stage) can be substantial in length and span large temperature
gradients. Wiring resistance can be kept low while maintaining low
thermal-conductivity by constructing the wires from a superconductor,
such as NbTi. 
\revisedtext{Since this wiring lies outside the tuned $LC\Rbolo$
  leg of the loop, its inductance $L$ can not be tuned out by the \fMUX\
  capacitor for more than one channel. }
This means it is necessary to keep the wiring inductance
low enough that the reactance is small compared to $\Rbolo$ at
the resonant frequency. Meeting this requirement is difficult with
traditional twisted pair wiring (50\,nH amounts to $0.3j\,\Omega$ at
1\,MHz, where $j$ denotes the imaginary component), 
so that low inductance microstrips are necessary for all or
part of this wiring (see \S\ref{s_cabling_enclosure_power}).

\subsubsection{Bandwidth and Detector Stability}
\label{s_stability}

To avoid under-damped or exponentially growing oscillations in the TES
response, the electrical power (half-width at half maximum, HWHM) 
bandwidth of 
the $LC\Rbolo$ filter $1/\tauLCR$ must exceed the TES thermal
bandwidth $1/\tauTES$ by a factor  
$1/\tauLCR \ge 5.8/\tauTES$~\cite{1998JAP....83.3978I}
(see Ref.~\cite{irwin_tes:2005} for a thorough description). 
\revisedtext{For the APEX-SZ and SPT detectors with $\Rbolo\simeq
  1~\Omega$ and thermalisation time constant $\tauweb \ge 5$\,ms
  (32~Hz roll-off frequency), we employ
   15.8~$\mu$H inductors to maximize the detector readout bandwidth
   while still meeting our constraint on cross-channel current
   leakage. This results in the requirement $\tauTES > 0.2$\,ms
   (0.8~kHz roll-off frequency). This allows for an order of magnitude
  range ($0.2$\,ms $< \tauTES < 5$\,ms) for the TES time constant. }
The TES time constants are increased to lie within this
range by adding a heat capacity in the form of an Au island that has a strong
thermal coupling to the TES. However, since the $\tauTES$ is sped up
by the loop-gain of the ETF, it is important to have a substantial
stability range. This stability requirement defines the minimum
allowed $LC\Rbolo$ bandwidth, and hence the channel spacing (described
below) for the \fMUX\ system.


\subsection{Multiplexer Channel Spacing}

The spacing between adjacent multiplexer channels is chosen such that
(1) Johnson noise leakage from adjacent channels is small, (2) bias
current leakage from adjacent channels does not spoil the detector
voltage bias, and (3) cross-talk between neighboring detectors on a
comb is small in comparison to the other sources of cross-talk in the
experiment. These conditions must hold for the $LC\Rbolo$ bandwidth
that is defined by the ETF stability requirements discussed in
\S\ref{s_stability}. For APEX-SZ and SPT a channel spacing of
75~kHz is used. The first two criteria are discussed below, while
cross-talk is discussed in \S\ref{s_cross_talk}.

Each bolometer in the multiplexer module is a source of broadband
Johnson noise. The SQUID input receives the sum of all
Johnson noise contributions from every channel, so it is necessary to
space the channels far enough apart in frequency so that the
Johnson noise from neighboring channels is attenuated to a negligible
level. This requirement is not the dominant constraint for channel
spacing. At a frequency of 75~kHz away from resonance, for $L=15.8\,\mu$H and
$\Rbolo=1\,\Omega$, the Johnson noise current of
 a detector is attenuated by
a factor 15, providing an increase of just 0.2\% to the total Johnson
current noise of its neighbor. The increase in total detector noise is
substantially smaller than this when other noise sources are also considered.
The Johnson noise leakage from spacings as close as 4 times the
$LC\Rbolo$ bandwidth (20~kHz, in this case) or less could be tolerated,
contributing a 3\% increase in the Johnson noise current of a
neighbor.


A more serious constraint on channel spacing is bias current leakage.
In addition to its own on-resonance bias, bolometer channel $i$ 
also sees a fraction of the current from the $i\pm 1$ nearest 
 carriers that neighbor in frequency space. The magnitude ratio of the 
off-resonance current $I^{\omega_{i\pm 1}}_{\mathrm{Ch}\,i}$ from the
neighboring carrier bias to the 
on-resonance current $I^{\omega_i}_{\mathrm{Ch}\,i}$ flowing through
bolometer channel $i$ is
\begin{equation} \label{e_carrier_leakage} 
\frac{ I^{\omega_{i\pm 1}}_{\mathrm{Ch}\,i} }{I^{\omega_i}_{\mathrm{Ch}\,i}} =
  \frac{\Rbolo}
          { \sqrt{ \Rbolo^2 + (\omega_{i\pm 1} L 
             - \frac{1}{ \omega_{i\pm 1} C_{\mathrm{Ch}\,i}} )^2 } } ,
\end{equation} 
where $\Rbolo$ is the bolometer resistance, $L$ is the series inductor, $C_{\mathrm{Ch}\,i}$ is the series
capacitor, and $\omega_{i\pm 1}$ is the angular frequency of the adjacent
bias carrier. For the channel spacing and parameters
considered here, this may be approximated 
$I^{\omega_{i\pm 1}}_{\mathrm{Ch}\,i} / I^{\omega_i}_{\mathrm{Ch}\,i} \approx
  \frac{\Rbolo}{2\Delta\omega L}$, where $\Delta\omega$ is the
channel spacing.

\revisedtext{
  Bias carrier leakage can be problematic, mostly because of cross-talk
  (discussed below), but also because the leakage is a quasi-current
  bias that provides a positive ETF contribution to the total bias power
  seen by the bolometer. While the negative ETF afforded by the
  on-resonance voltage bias will dominate, the contribution from
  carrier leakage can cause instability and will also enhance the
  bolometer responsivity slightly.}


All of these effects are
rather small in practice. For a spacing of 50~kHz (10
times the $LC\Rbolo$ bandwidth), the electrical bias power from one of
the nearest neighbors is roughly 1\% of the on-resonance power
assuming that all bolometers are biased at the same voltage amplitude.
For channels biased near the center of the bias-frequency comb, the
total off-resonance bias power is approximately 3\%. This leads to a
3\% increase in both the responsitivity and the effective time
constant, and a 3\% decrease in the maximum loopgain. At our spacing of
75~kHz (15 times the $LC\Rbolo$ bandwidth), these effects are reduced to
1\%.

%


\subsection{Cross-talk} 
\label{s_cross_talk}

The multiplexer contributes channel-to-channel cross-talk due to (1)
inductor cross-coupling, (2) bias carrier leakage, and (3) signals
from a detector causing heating in neighbor detectors due to the
voltage drop across non-zero \squid\ or wiring impedance.
Contributions from these mechanisms, discussed below, are smaller than
the $\sim$1\% optical cross-talk that typically exists between
neighboring pixels for APEX-SZ and SPT.

(1)  Inductor cross-coupling: 
The mutual inductance $M_{i,j}=k_{i,j} L_i L_j$ between inductors $L_i$ and
$L_j$ of two channels $i$ and $j$ in a multiplexer
module creates cross-talk. The carrier current $I_i$ flowing
through the $i^{th}$ bolometer channel induces a voltage 
$|V_{j}|=\omega_i M_{i,j} I_i$ in the inductor $j$ that physically
neighbors channel $i$. As $\Rbolo^{i}$ changes in response to sky
signals, the voltage $V_{j}$ is modulated as well. This modulation
produces small changes in the current flowing in channel $j$
since $\Rbolo^{j}$ forms a small fraction of the total $LC\Rbolo^{j}$
impedance at frequencies far from its resonance. The measured coupling
coefficient~\cite{2005ApPhL..86k2511L} between neighboring inductors
on the same chip is $k=0.010\pm0.002$ for the \fMUX\ system. This form
of cross-talk can be made negligible by ensuring that channels with
physically neighboring inductors are not neighbors in frequency space
(i.e., that physically neighboring inductors in this implementation
have carrier bias frequencies separated by more than 150~kHz).

(2) Bias carrier leakage: As $\Rbolo$ changes in response to sky
signals incident on bolometer $i$, both its on-resonance carrier at
frequency $\omega_i$ and
the off-resonance leakage bias carriers at neighboring frequencies
$\omega_{i\pm 1}$ are modulated, creating
cross-talk. This mechanism encodes a cross-talk sky-signal on the
leakage current waveform.  
 The signal appears as an amplitude modulation in the on-resonance
  carrier, but leads to modulation in both the amplitude and the phase
  of the off-resonance carrier.  Thus a full quadrature demodulator
  would see cross talk signal in both the I- and Q-components of the
  demodulator output.  However, the demodulator measures only the
  I-component modulations. For an off-resonance $LC\Rbolo$ channel
  $i\pm 1$ at carrier frequency $\omega_{i}$, the amplitude
  fluctuation is:
\begin{eqnarray}
I^{\omega_i}_{\mathrm{Ch}\,i\pm 1} && = \frac{V_\mathrm{bias}^{\omega_i}}
          { \Rbolo + j \omega_i L + 1/(j \omega_i C_{\mathrm{Ch}\,i\pm 1}) } \\
       && \simeq \frac{V_\mathrm{bias}^{\omega_i}} { \Rbolo + j 2\Delta \omega\, L } \\
       && \simeq  \frac{V_\mathrm{bias}^{\omega_i}}{j 2\Delta \omega\, L}
       \left( 1 + \frac{j\Rbolo}{2\Delta \omega\, L} \right) ,
\end{eqnarray}
where a first-order Taylor expansion in the parameter 
$\frac{ j \Rbolo} { 2\Delta \omega\, L }$ has been performed for the
last equality.  
\mynewtext{This results in a current modulation
with a change in bolometer resistance of}
\begin{equation} \label{e_modulation_off}
  \frac{ \Delta I^{\omega_i}_{\mathrm{Ch}\,i\pm 1} }{ \Delta \Rbolo } \simeq
  \frac{V_\mathrm{bias}^{\omega_i} } { (2\Delta \omega\, L )^2 } ,
\end{equation}
\mynewtext{which should be compared to the
current modulation }
\begin{equation} \label{e_modulation_on}
  \frac{ \Delta I^{\omega_i}_{\mathrm{Ch}\,i} }{ \Delta \Rbolo } \simeq
  \frac{-V_\mathrm{bias}^{\omega_i} } { \Rbolo^2 } 
\end{equation} 
that would occur on-resonance.
The magnitude ratio of Equation~\ref{e_modulation_off} to
Equation~\ref{e_modulation_on}
\begin{equation}
\left| \frac{ \Rbolo^2 } { (2\Delta \omega\, L )^2 } \right| 
\end{equation} 
is a good approximation for this cross-talk. This is equivalent to
the square of the current leakage ratio presented in Equation~\ref{e_carrier_leakage}. 
For our
circuit parameters
($\Rbolo=0.75\,\Omega,~\Delta\omega= 2\pi \cdot 75\,\mbox{kHz},~L=15.8\,\mu\mbox{H}$),
this effect results in $\sim$0.25\% cross-talk. Note that this is larger
than the cross-talk \revisedtext{estimate} that was derived in
Ref.~\cite{2005ApPhL..86k2511L}, which did not 
\revisedtext{correctly} take into account the
phase of this cross-talk signal.

\begin{figure}
  \includegraphics[width=0.5\textwidth,clip=]{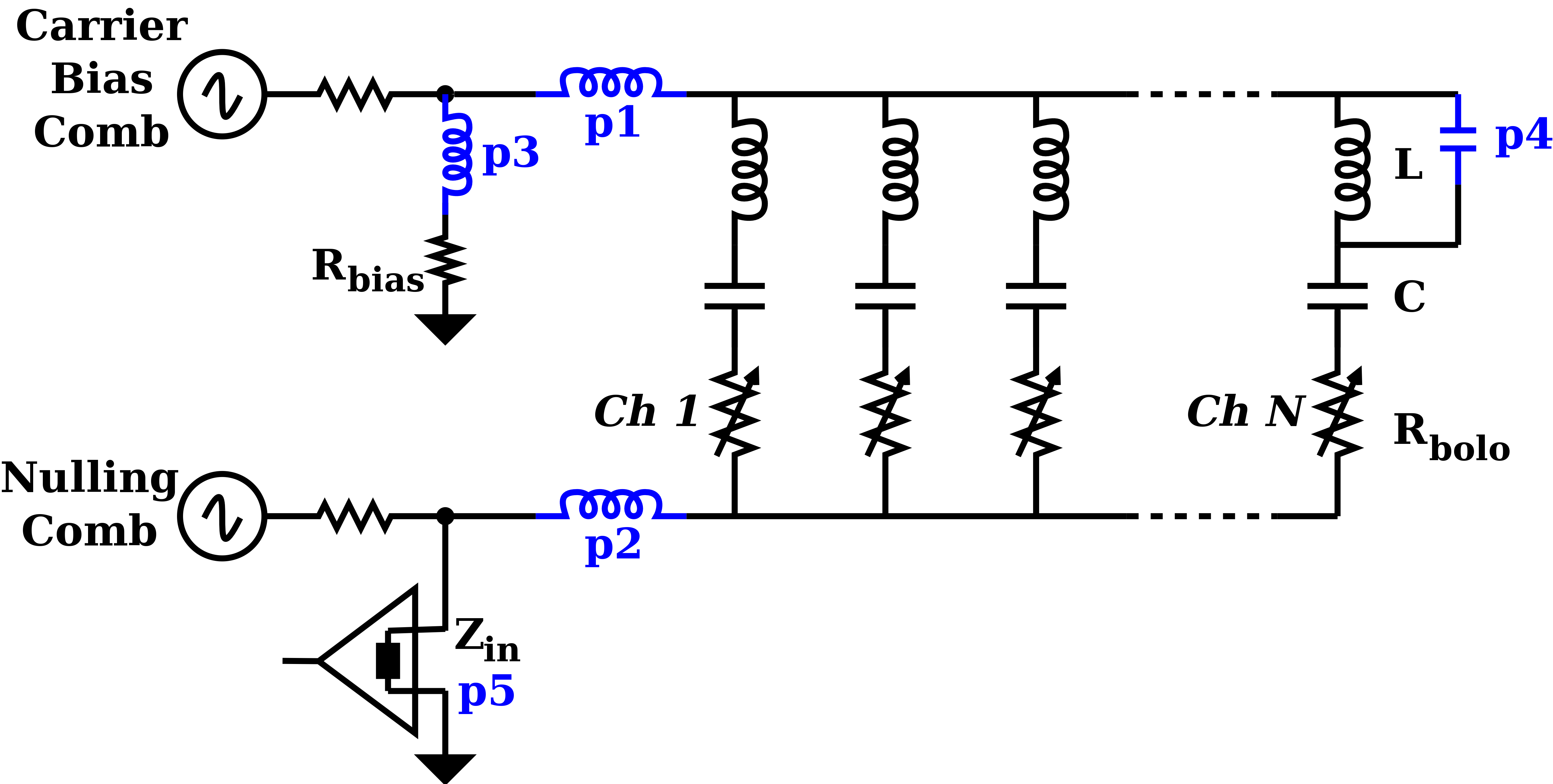}
  \caption{ \label{f_mux_strays} 
  Stray reactances in the sub-kelvin multiplexer circuit are
  problematic since the circuit operates at MHz frequency with
  characteristic impedances $\sim 1 \Omega$. The stray inductance of
  the wiring connecting the bias resistor ($R_\mathrm{bias}$) to the
  multiplexer circuit is shown at p1 and p2, the stray inductance in
  series with $R_\mathrm{bias}$ is shown at p3, the stray
  capacitance in parallel with a channel's inductor is shown at 
  p4, and the non-zero input impedance of the transimpedance
  pre-amplifier is shown at p5. }
\end{figure}

(3) Non-zero \squid/wiring impedance: A change in bolometer impedance
of channel $i$ in response to sky-signals, causes the voltage at
carrier frequency $\omega_i$ across the stray \squid\ and wiring
reactance to be modulated, which in turn causes a modulation in the
leakage current power deposited in neighboring detectors $i\pm 1$.
This mechanism transmits cross-talk power from the on-resonance
detector to the off-resonance neighbors, which is the opposite
transfer direction compared to mechanism (2).
The voltage
$V_\mathrm{module}$ across the sub-kelvin multiplexer module (measured
from the right side of p1 to the right side of p2 in
Figure~\ref{f_mux_strays}) is not exactly equal to the voltage
$V_\mathrm{bias}$ across the bias resistor. 
This discrepancy arises from the non-zero complex input impedance
of the \squid\ system (p5 in Figure~\ref{f_mux_strays}) and
the stray inductance of the wiring (p1 and p2 in
Figure~\ref{f_mux_strays}) that connects the module to the \squid\
input (together labelled $V_\mathrm{stray}$) such that
$V_\mathrm{module} =V_\mathrm{bias} - V_\mathrm{stray}$. For the
purposes of this discussion, the p3 and p4 stray components in
Figure~\ref{f_mux_strays} can be ignored. 
%
Consider the current $I^{\omega^i}_{\mathrm{Ch}\,i}$ at carrier frequency $\omega_{i}$ that flows
predominantly through the $i^{th}$ bolometer. The change in voltage
across the module due to a change $d I^{\omega^i}_{Ch\,i}$ is
\begin{eqnarray}
d V_\mathrm{module} && = d V_\mathrm{bias} -d V_\mathrm{stray} \\
                           && = -d V_\mathrm{stray} = - d I^{\omega^i}_{\mathrm{Ch}\,i}  Z_\mathrm{stray} \\
                           && \simeq - d I^{\omega^i}_{\mathrm{Ch}\,i} \cdot j \omega_i L_\mathrm{stray}
\end{eqnarray}
at fixed $V_\mathrm{bias}$. We have assumed that the \squid\ system and wiring impedances
$Z_{stray}$ are dominated by an inductive term $L_\mathrm{stray}$.
This voltage induces a leakage current $d I^{\omega^i}_{\mathrm{Ch}\,i\pm1}$
through the neighboring $i\pm 1$ bolometer channel
\begin{equation}
  d I^{\omega^i}_{\mathrm{Ch}\,i\pm1} = \frac{ d V_\mathrm{module} }{ Z^\mathrm{LCR}_{\mathrm{Ch}\,i\pm1} }
  \simeq \frac{- d I^{\omega^i}_{\mathrm{Ch}\,i} \cdot j \omega_i L_\mathrm{stray} }
             { j2\Delta \omega  L } ,
\end{equation}
where $Z^\mathrm{LCR}_{\mathrm{Ch}\,i\pm 1} \simeq j2\Delta \omega L$ is the 
impedance 
of the neighboring $i\pm 1$ $LC\Rbolo$ leg of the cold multiplexer
module at frequency  $\omega_{i}$. This current fluctuation deposits
Joule heating power 
\begin{equation}
  d P^{\omega^i}_{\mathrm{Ch}\,i\pm 1} \simeq 2 \Rbolo \, 
            I^{\omega^i}_{\mathrm{Ch}\,i\pm 1} \cdot d I^{\omega^i}_{\mathrm{Ch}\,i\pm 1}
\end{equation}
in the neighboring bolometer channel, which should be compared to the
signal in Channel $i$, 
\begin{equation} 
  d P^{\omega^i}_{\mathrm{Ch}\,i} \simeq d I^{\omega^i}_{\mathrm{Ch}\,i} \, V_\mathrm{module} .
\end{equation}  
The ratio of these power fluctuations is the cross talk,
\begin{equation} 
\frac { d P^{\omega^i}_{\mathrm{Ch}\,i\pm 1} } {d P^{\omega^i}_{\mathrm{Ch}\,i} }
  \simeq -
  \frac{ I^{\omega^i}_{\mathrm{Ch}\,i\pm 1} }{ I^{\omega^i}_{\mathrm{Ch}\,i} } \,
  \frac{ \omega_i }{\Delta\omega} \,
  \frac{ L_\mathrm{stray} }{ L } .
\end{equation}
For the \fMUX\ system with $L  / L_\mathrm{stray} \sim 150$, 
$\Rbolo=0.75\,\Omega$,
$ \Delta \omega = 2\pi \cdot 75$\,kHz,
and taking $\omega_i=2\pi \cdot 750$\,kHz, 
this cross-talk amounts to 0.3\%.

While the three forms of readout system cross-talk discussed above would
not be present in a non-multiplexed system, we note that a dominant
form of electronic cross-talk for a non-multiplexed system can be
eliminated for the \fMUX\ system. Capacitive coupling of wires within
multi-wire cables that carry signals from different detector channels
is typically $\sim$1\% for bundles of twisted pairs. For a
non-multiplexed system, all detector signals normally occupy the same
frequency bandwidth as they are transmitted through these cables,
resulting in 1\% cross talk with neighbors. For the \fMUX\ system,
each wire carries a module of detectors, each at a unique bias
frequency. Since the $\sim$100~Hz signal bandwidth is a small fraction
of the total bandwidth, the bias frequencies of modules that are
bundled together can be chosen to be unique. This places the cable
cross-talk signals in a portion of the bandwidth that is discarded by
the signal demodulation. This means that, overall, this multiplexed
readout system has lower cross-talk than a typical 
non-multiplexed system.

\subsection{Channel Count}

Considering the results of the previous sections, the number of 
channels that can be multiplexed together without adversely affecting 
the noise or cross-talk for a system with a given set of parameters is
derived as follows.

The $LC\Rbolo$ HWHM power bandwidth is chosen to be at least 5.8 times
wider than the inverse of the fastest expected TES time constant,
$1/(2\pi\tauTES)$ (see \S~\ref{s_stability}). This has very
conservatively been chosen to be 5~kHz in this system, allowing an
order of magnitude range of $0.2$\,ms $< \tauTES < 5$\,ms so that the
time constants are ordered appropriately as $5.8\,\tauLCR < \tauTES <
\tauweb$.

The spacing between $LC\Rbolo$ resonances is chosen so that the total
bias carrier leakage current is less than 20\% of the on-resonance
carrier current, and the cross-talk is less than 0.5\%. For the
parameters of this system, this results in a spacing of 15 times the
$LC\Rbolo$ bandwidth, 75~kHz. This spacing also keeps Johnson noise
leakage at a negligible level.

Having specified the channel spacing, the number of channels is
defined by the bandwidth of the low impedance amplifier. The \squid\
shunt-feedback circuit (discussed in \S\ref{s_squid}) for this system
has a bandwidth of 1~MHz. The carriers are positioned between about
400~kHz and 900~kHz, ensuring odd harmonics from low-frequency
carriers do not fall within the bandwidth of higher frequency channels.
Odd harmonics are the dominant form of distortion from \squid\
non-linearity. Bias carriers below 400~kHz are avoided because of
the difficulty of finding high-Q capacitors with sufficiently high capacitance
that behave well at cryogenic temperatures. These considerations
result in 7~channels per multiplexer module so that the number of
wires connecting the sub-kelvin multiplexer to the 4~K \squid\ is the
total number of detectors multiplied by 2/7.

The per-module channel count for APEX-SZ and SPT is conservative. A
small increase in channel count could be achieved by slowing the TES time
constants $\tauTES$ so that they are better matched to the absorber
thermalization time $\tauweb$, allowing narrower channel spacing.
Substantial increases could be achieved by implementing a \squid\ 
flux-locked loop (FLL) with substantially larger bandwidth such as
Ref.~\cite{2009arXiv0901.1919L}. Finally, much narrower
channel spacing could be realized by using a sharper band-pass filter
than the single pole $LC\Rbolo$ employed here, although this would
substantially complicate the cold circuit.

\subsection{Sub-kelvin Multiplexer Implementation}
\label{s_cold_mux_implementation}

Each sub-kelvin multiplexer module consists of a custom-fabricated
chip of seven inductors, seven commercially available ceramic
capacitors with negative-positive-zero (NP0)
dielectric material, and a custom printed circuit
board (shared by all multiplexer modules for a particular detector
wafer) on which these components are mounted. 
The multiplexer modules are
interfaced with aluminum wire bonds to the detector wafer as shown in
Figure~\ref{f_bolometer_photo}. The circuit board is thermally sunk to
the same 250\,mK temperature stage as the bolometers.

The inductors were custom fabricated at the 
Northrop-Grumman superconducting micro-fabrication facility
with a lithographic process. Each chip
consists of eight inductors, of which we use seven. One inductor is
about 1.5\,mm\,$\times$\,1.5\,mm, consisting of a 140\,turn spiral
(2\,$\mu$m width niobium trace with pitch of 4\,$\mu$m) separated by a
50\,nm SiO$_2$ insulating layer from a niobium, flux-focusing washer
that is left floating. The 15.8\,$\mu$H inductance is the same for all
channels in the module. 
The Northrop-Grumman superconducting microlab
is no longer available, and inductors for new implementations of the
\fMUX\ system are now being
fabricated at the National Institute for Standards and Technology
(NIST).


The frequency of each channel resonance is set by the
series capacitor, for which we use commercially available surface
mount capacitors. Our target spacing for
the 7-channel multiplexer is 75\,kHz. Several capacitors, pre-selected
by measuring their capacitances individually, are stacked in parallel
to achieve the target capacitance for each channel. The 
capacitances change by several percent when cooled from room
temperature to 0.25\,K, but in general all devices shift
proportionately and the channel spacing is sufficiently preserved.

\begin{figure} \includegraphics[width=0.5\textwidth,clip=]{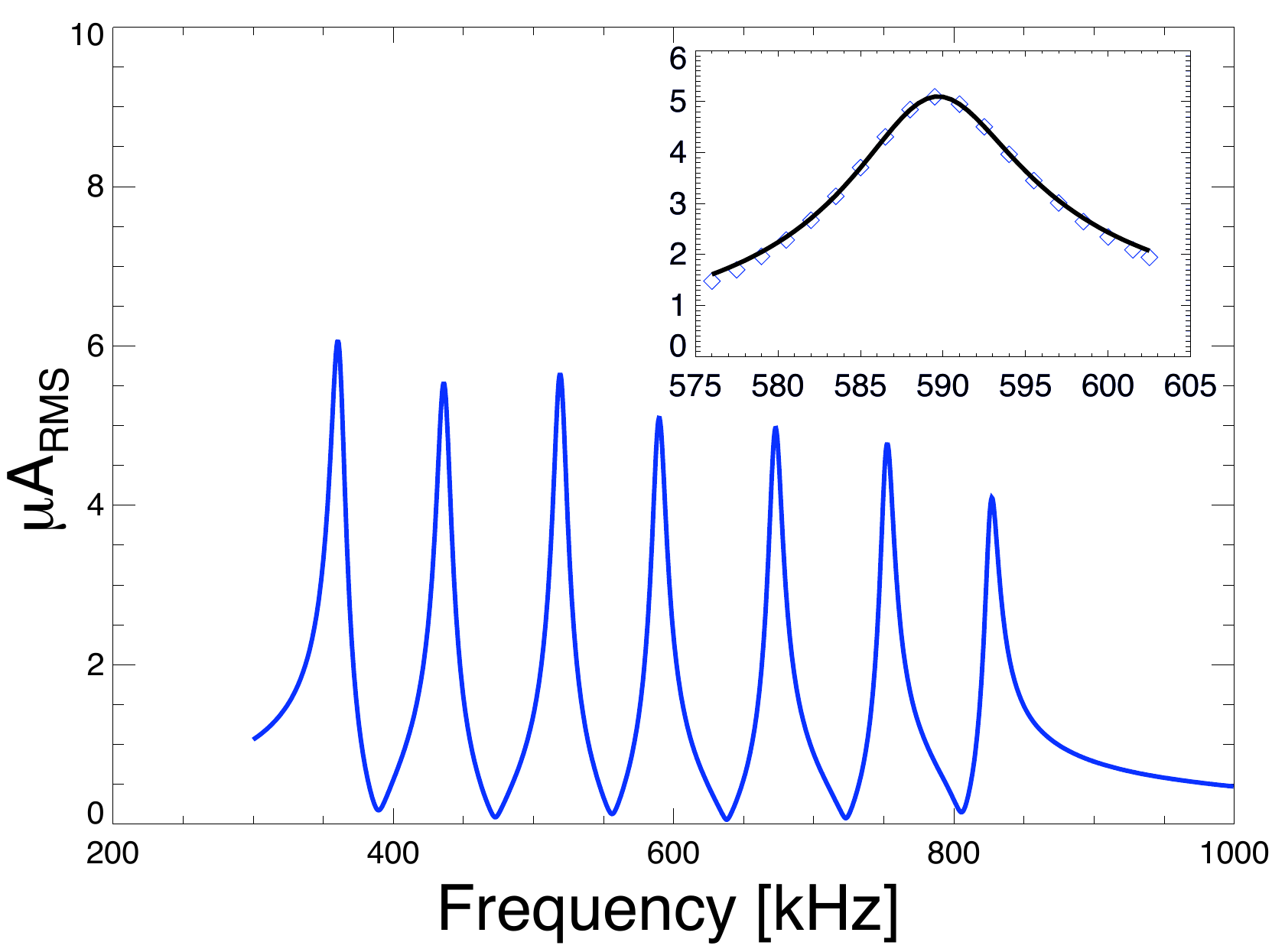}
  \caption{ \label{f_netanal} The resonant frequencies for each
    $LC\Rbolo$ leg of the multiplexer circuit are measured by
    frequency-sweeping a small amplitude bias carrier across the range
    of carrier frequencies with the bolometer stage temperature held
    just above the detector superconducting transition. The
    resultant network analysis data, shown in the amplitude vs.
    frequency plot above, is fitted with a model to determine the
    optimum carrier bias frequency for each bolometer. One such fit is
   shown in the inset.}
\end{figure}

The resonant frequencies for each multiplexer module are measured by
frequency-sweeping a small amplitude bias carrier across the range of
carrier frequencies with the bolometer stage temperature held just
above the detector superconducting transition. The amplitude of the
current through the \squid\ is measured. A representative `network
analysis' amplitude vs. frequency plot for a multiplexer module is
shown in Figure~\ref{f_netanal}. Each peak represents one
bolometer channel, with its location, width, and height defined by the
$LC\Rbolo$ parameters. The bias frequency for each bolometer is
determined by fitting a model to each peak that includes parameters to
account for the current flowing through off-resonant $LC\Rbolo$ legs
in the multiplexer module.

\subsection{Effects of Stray Reactances} \label{s_cold_strays} 

Although the sub-kelvin multiplexer components are relatively simple,
non-idealities in the form of stray inductances and
capacitances can be problematic, since the circuit operates at MHz
frequencies with characteristic impedances of about an ohm. The effect
of these strays are discussed below.

The inductance of the cryogenic wiring that connects the bias resistor
($R_\mathrm{bias}$) at 4\,K to the sub-kelvin $LC\Rbolo$ multiplexer
circuit is the most important stray, shown at p1 and p2 in
Figure~\ref{f_mux_strays}. This inductance is outside each
$LC\Rbolo$ resonant circuit, and so cannot be tuned out at
all of the carrier resonant frequencies. 
The voltage drop across this reactance
reduces the voltage across the bolometer and partially spoils the
constant voltage bias that is necessary for electro-thermal feedback. The effect is
worse at higher frequency. To control this stray component, the wiring is
constructed of two segments: a low inductance 
($\sim$0.3\,nH/cm) broadside coupled
stripline is used for the majority of the wiring distance, followed by a
short segment of NbTi twisted pair 
($\sim$7\,nH/cm)  that is necessary to create a
low thermal conductivity gap between the two temperature stages.
The total inductance for this wiring in the
system implementation is about 105\,nH, amounting to 
(0.2-0.6)$\,j\Omega$ at 300-900\,kHz,
increasing the impedance of the on-resonance 0.75\,$\Omega$ bolometer
circuit ($\Rbolo + j\omega L_\mathrm{stray}$) by 3-27\%.

There is also a much smaller stray inductance in series with
$R_\mathrm{bias}$ (p3 in Figure~\ref{f_mux_strays}) that has the
opposite effect compared to the above, increasing the bias
voltage across the bolometer. The circuit board housing the \squid
s and 30\,m$\Omega$ bias resistors has been optimized to minimize this
stray, which is estimated at $\sim$4\,nH. The total $R_{bias} + j
\omega L_\mathrm{stray}$ reactance increases from 31~m$\Omega$ at
300\,kHz to 38\,m$\Omega$ at 900\,kHz. While this stray results in a
large change (as much as 26\% at 900\,kHz) in the effective bias
voltage across the bolometer, this phase and amplitude modified
voltage still provides an effective bias for the purposes of ETF. To
see this, we note that it would be possible to bias the bolometer using
just an inductance in place of the bias resistor, eliminating Johnson
noise from this component. The drawback is that the strong dependence
on frequency of this inductance-bias voltage strategy would complicate
the setup and tuning of the system. An inductive or capacitive voltage
divider, as described in Ref.~\cite{2009AIPC.1185..245V}, would address
both issues.

Both stray inductances described above affect the phase of the bias
carriers in addition to their amplitude. Fortunately, the phase shifts
of the two effects partially cancel.  However, because of the lack of
phase tuning capability for the demodulator in the backend electronics
in this implementation of the readout (\S\ref{s_demodulator}), this
phase shift contributes to a slight degradation of noise at high
frequency.

A stray capacitance is in parallel with the inductor (p4 in
Figure~\ref{f_mux_strays}) that forms the $LC\Rbolo$
series resonance for each multiplexer channel. This is due to a flux-focusing washer 
below the spiral trace of each inductor. With the present architecture, the
resonance is in the 16\,MHz range. 
While this feature is well above
the multiplexer bandwidth, the phase shift it produces can cause
stability problems for the \squid\ pre-amplifier feedback, discussed
in \S\ref{s_squid}.

The transimpedance pre-amplifier (a \squid\ array in this
implementation) has non-zero input reactance (p5 in
Figure~\ref{f_mux_strays}). 
This means that currents injected along the nulling wire
are split between two return paths: 
(1) the desired path through the pre-amplifier, and (2) through the
bolometers. We refer to path (2) as ``nuller leakage''. Like the
carrier voltage across the bias resistor, this leakage also appears as
a voltage across the bolometers, with the pre-amplifier reactance
playing the role of a bias resistor. This effect is more pronounced at
higher frequencies. When the nuller current phase and amplitude is
adjusted so that the net (carrier and nuller) current through the
pre-ampifier is zero, the nuller current will be an exact inverted
copy of the current through the bolometer. 
\mynewtext{
The nuller current is
adjusted in this manner when the bolometers are tuned, forming a
`virtual ground' at the pre-amplifier input so that the net voltage
across the input is zero after each tuning.
Changes in observation loading conditions between bolometer tunings will
result in a non-zero voltage across the pre-amplifier and alter the
voltage bias across the bolometer.}
%


\mynewtext{
Achieving the design requirements outlined in
\S~\ref{s_design_requirements} has required careful handling of the
stray components described above. Future readout systems that employ
higher frequency carriers to multiplex a larger number of bolometer
channels will need more advanced techniques to correct for these stray
components.  }
\section{SQUID Pre-amplifier and Room Temperature Electronics}
\label{s_squid_and_electronics}

Having described the design considerations for the sub-kelvin
components in the previous section, we now describe
the \squid\ control and backend
electronics in this section, beginning with the
generation of the carriers that bias the detectors, then describing
 the \squid\ pre-amplifier and its room temperature
control electronics, and ending with the demodulation electronics.

The topology of the system is shown in
Figure~\ref{f_fmux_implementation} for one 7-channel \fMUX\ module.
APEX-SZ uses 40 \fMUX\ modules and SPT uses 120. Seven bias carriers
are synthesized separately using Direct Digital Synthesizers (DDS)
located on room temperature oscillator/demodulator boards. These
analog carriers are added together to form a bias `comb' and
transmitted via a twisted pair cable and the room temperature
\squid\ controller board to the cryostat, where a 30\,m$\Omega$ bias
resistor at 4~K produces a voltage bias across the bolometers
operating at 250~mK. LC resonant filters in series with the bolometers
select a single carrier tone for each bolometer. Changes in optical
power on the detectors induce amplitude modulation of the carrier currents which
are summed together and pass through the input of a series-array
\squid\ device. The \squid\ is coupled to a room temperature operational amplifier to 
form a shunt-feedback flux-locked-loop circuit. A separate `nulling
comb', which is simply an inverted version of the carrier comb, is
transmitted from the oscillator/demodulator boards to the \squid\
input. This serves to cancel the large carriers, reducing the dynamic
range requirement for the \squid. The nulled and amplified comb is
transmitted via a cable to a bank of seven demodulators located on the
oscillator/demodulator boards. Each amplitude-modulated carrier is
demodulated separately, low-pass filtered, and digitized. The
digitized sky-signals are transmitted to a data acquisition computer
and recorded on disk. Each of these components is described in detail
below.

\subsection{Carrier Generation} \label{s_carrier_gen}

\begin{figure}
\includegraphics[width=0.45\textwidth,clip=]{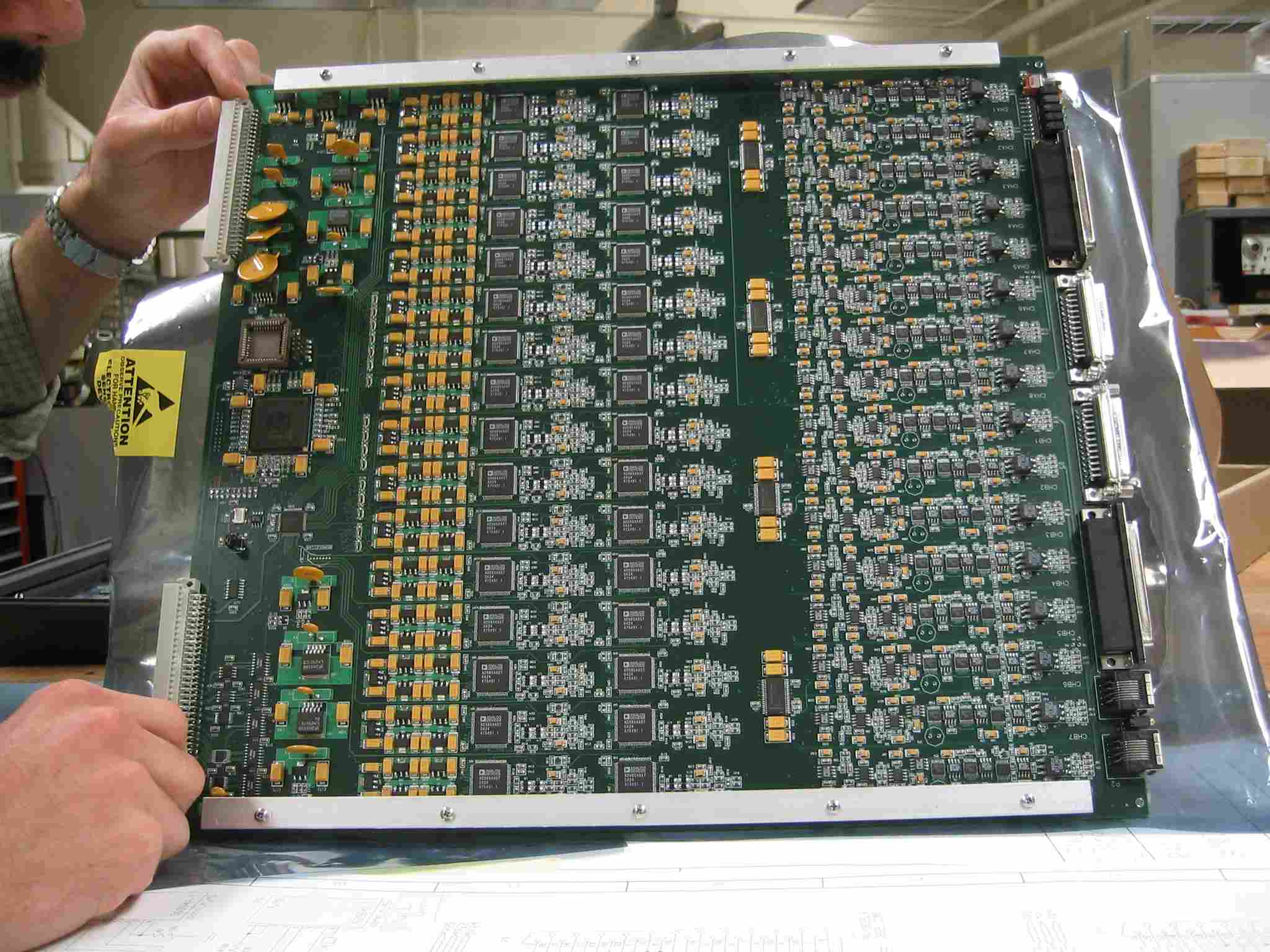}
\caption{ \label{f_oscdemod_photo} (Color online) Photo showing the 
  oscillator/demodulator circuit board that provides the 
  biases and demodulates the outputs for 14~bolometer channels.}
\end{figure}

\begin{figure}
 \includegraphics[width=0.5\textwidth,clip=]{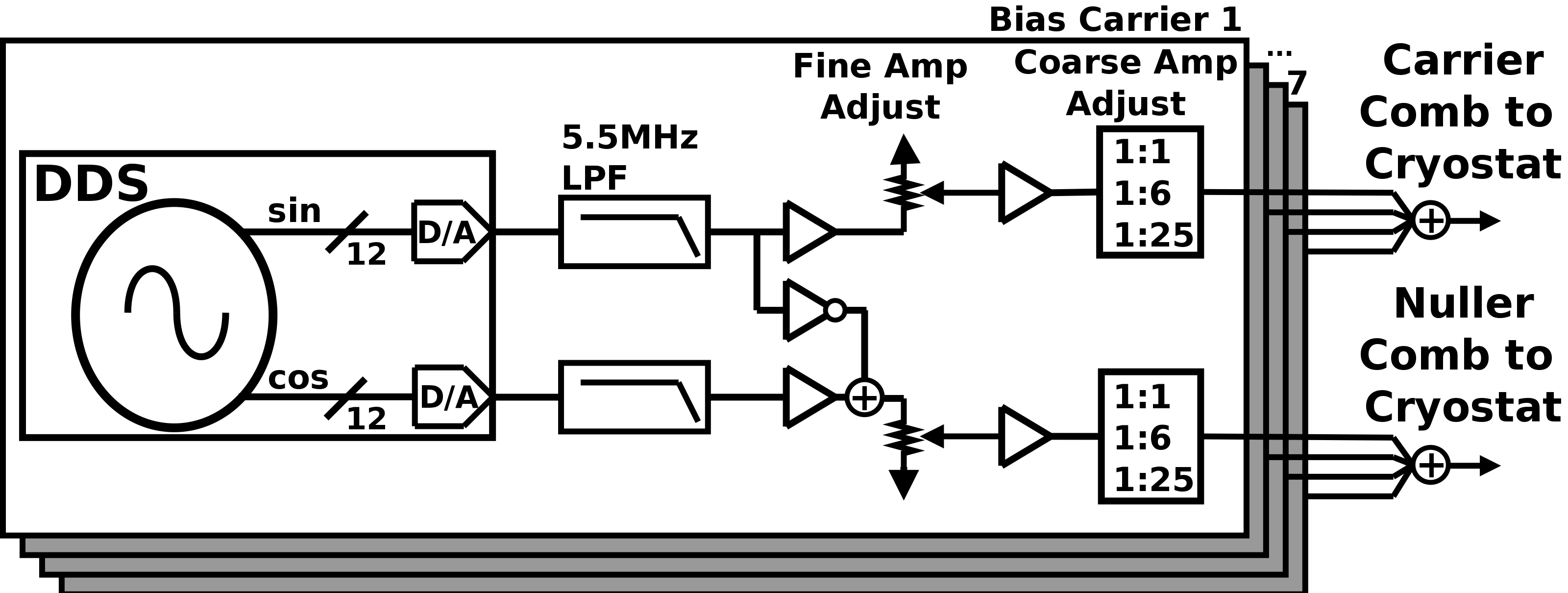}
 \caption{ \label{f_carrier_generation} Block diagram showing the carrier and 
                         nuller generation circuits. 
The sine and cosine outputs at the DDS have amplitude adjustment
registers that are not shown. The DDS amplitudes are used to control
the phase of the nuller signals by adjusting the admixtures of sine
and cosine waves.}
\end{figure}

The bolometer bias carriers are synthesized on custom circuit boards
(Figure \ref {f_oscdemod_photo}) that each handle 14~bolometer
channels.
In addition to synthesizing the carriers, these boards also perform
the demodulation and digitization of the bolometer outputs described
in \S\ref{s_demodulator}. The layout of these boards is
optimized to minimize cross-talk between channels and pickup from
other electronics systems. A block diagram showing the circuit used to
generate the carrier comb and nuller comb is shown in
Figure~\ref{f_carrier_generation}.

The bias for each bolometer is generated with a Direct Digital
Synthesizer (DDS, Analog Devices AD9854) that produces a sine wave by
storing the waveform instantaneous phase in an accumulator and
incrementing this accumulator every clock cycle. It uses the phase to
address a corresponding amplitude in a look-up table, and sends that
amplitude to a digital-to-analog converter (DAC) sampling at 40\,MHz.
Because the phase information is held in only one register, it is not
costly to achieve many bits of frequency resolution (these devices use
a 48-bit phase accumulator and have $\mu$Hz resolution). Having
accurate amplitude information is more costly since each memory
location within the look-up table needs this accuracy. Our devices
provide 12-bit amplitude information. The output waveform can be
multiplied pre-digitization by an amplitude adjustment factor that
ranges from 0 to 1. In addition to the DAC providing the sine wave
output, each DDS has a second DAC output that has a fixed 90 degree
phase offset (i.e., a cosine wave) and separate amplitude adjustment
factor. The DDS chips are the single biggest power consumers in the
readout system, each one consuming $\sim$0.4~W.

The DDS output signals are differential and each pass through low-pass
5.5~MHz analog filters to reduce clock bleed-through and limit the
noise bandwidth of the bias carriers.
The sine-wave output is used for the bolometer bias signals. An 8-bit
digital potentiometer is used to control this amplitude. A coarse
voltage attenuator can be enabled to provide further amplitude
adjustment, allowing for good noise performance over a broad range of
bolometer parameters. The analog differential output currents from
eight oscillator circuits are summed together to produce a bias comb
that is transmitted to the cryostat via a twisted pair cable and
the \squid\ controller boards. Seven of these eight outputs are used
to bias the bolometers, the eighth is used as a test signal that can
be injected one at a time to each of the bolometers in the comb.

The full scale carrier to white noise ratio is $\approx 6\times 10^6$\rtHz, 
dominated by the digitization noise of the DDS. More important than
the white noise level however, is the carrier sideband noise, since
this noise eventually dominates at low frequencies and limits our
ability to observe large angular scale signals on the sky. Carrier
sideband noise can originate from clock jitter, DDS voltage reference
jitter, and/or current noise in the transistors used in the DDS DAC.
This system is not sensitive to clock-jitter, because the demodulator
is driven by the same oscillator as the bias, approximately canceling
the effect. The dominant source of sideband noise is the low frequency
noise of the DAC output transistors. This noise is proportional to the
amplitude $A_c$ of the carrier and measured to be $\sim A_c \times
10^{-5}$/\rtHz\ at 1~Hz away from the carrier, with a 1/f power
spectrum.

In addition to the bias carrier generation, the same DDS circuits are
used to provide the nulling signal that zeros the
carrier amplitude just before the \squid\ input. The nuller consists
of an inverted version of the bias sine-wave, with a small admixture of
the cosine signal from the second DDS output. The sum of these two
signals is also a sine wave with an phase that can be programmed (by
adjusting the cosine amplitude) from about 158 to 202~degrees from the
bias carrier. This phase adjustment compensates for 
analog phase shifts between the carrier and nuller that arise because
the two signals take different paths through the cold multiplexer
(Figure~\ref{f_fmux_implementation}). The filtering, amplitude
adjustment, coarse attenuation, and summing circuits are the same for
the nulling signals as for the carriers.

This scheme for synthesizing the nulling signal was motivated so that,
when the cosine admixture is small, the bias low-frequency carrier
sideband noise would be inverted for the nuller signal and subtract
to zero at the input of the \squid . This sideband-noise
nulling was demonstrated to work very well when adding the carrier to
the nuller through a resistor. However, the scheme does not work when
used to bias bolometers, since the bolometer has an effective negative
resistance within its active bandwidth. This means the sideband noise
of the carrier is inverted by the bolometer, while its MHz carrier is
not inverted. The net result is that the carrier sideband noise adds
coherently to the nuller sideband noise. Though this results in excess
low-frequency readout noise for this system, other sources of
low-frequency noise dominate our detector noise (e.g., residual
atmosphere, temperature drifts, etc.), and the low frequency noise
from the readout does not dominate the
overall low frequency noise of the experiment.
The digital \fMUX\ system described in
Ref.~\cite{2008ITNS...55...21D}) uses a different scheme for nulling
that avoids this excess readout noise. 

\subsection{\squid\ Electronics} \label{s_squid}

\begin{figure}
\includegraphics[width=0.475\textwidth,clip=]{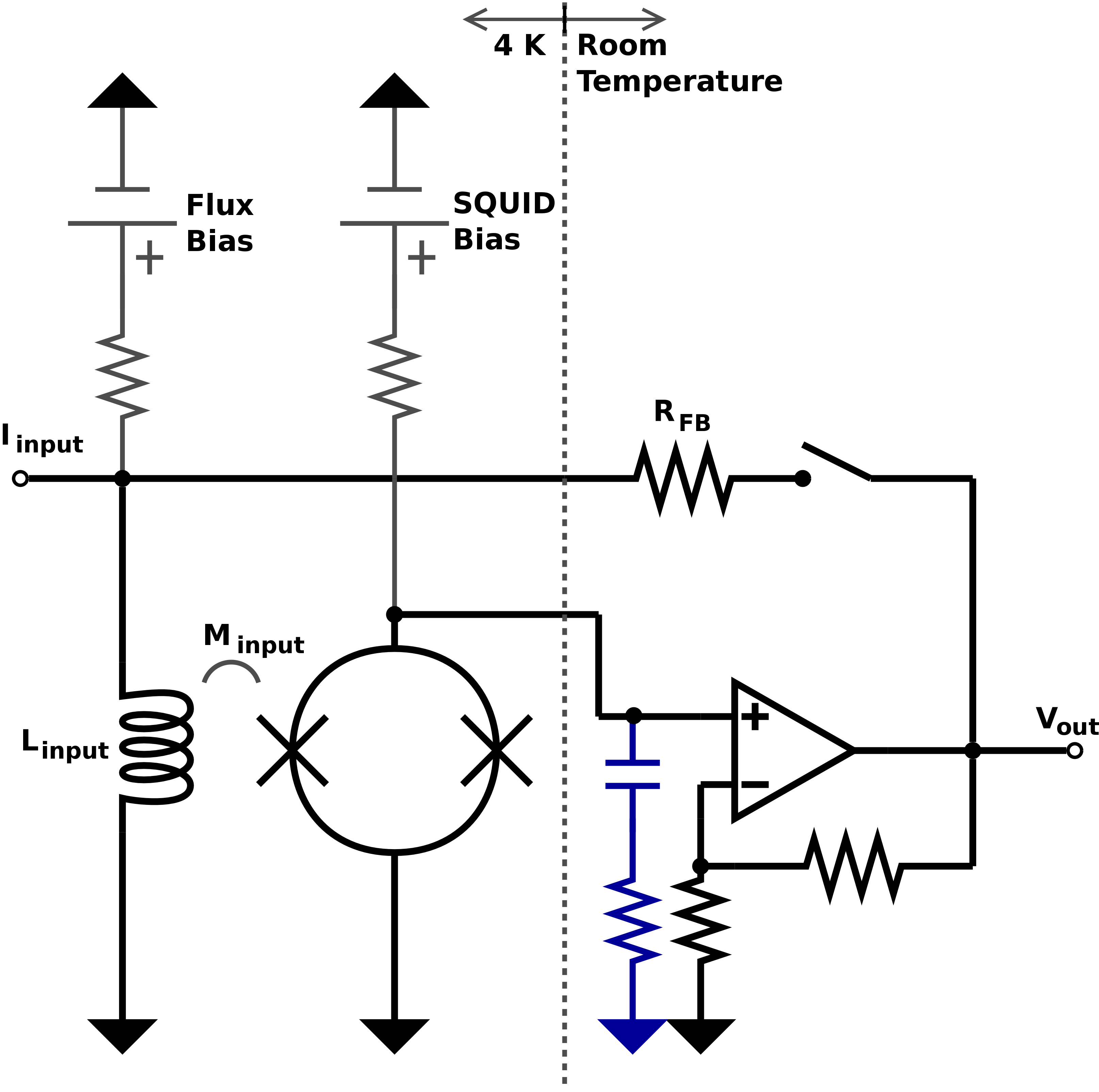}
\caption{ \label{f_SQUID_FLL}
The \squid\ shunt-feedback circuit is shown.
The capacitor and resistor that together form the lead-lag filter
(refer to text) connect the non-inverting input of the operational amplifier to ground. 
}
\end{figure}

Transimpedance amplification of the bolometer currents is achieved
with a series-array \squid\ operating in a shunt-feedback circuit with
a room temperature op-amp as shown in Figure~\ref{f_SQUID_FLL}. This
circuit has the following properties: (1) Its input impedance is sufficiently 
low so as not to spoil the voltage bias across the bolometers. 
(2) It provides sufficient transimpedance so that the output signals
can be interfaced to standard room-temperature amplifiers. (3) Its
noise, referred to the input, is small compared to the bolometer
noise. (4) It is sufficiently linear that intermodulation distortion
is not an issue for the system operation.

For the \fMUX\ system, the \squid s are housed in groups of eight on
custom \squid-mounting boards that are heat-sunk to the 4~K cryostat
mainplate. The rest of the shunt-feedback circuit resides on custom
room temperature \squid-controller circuit boards, each of which
manages eight \squid s. In addition to completing the shunt-feedback
circuit, these electronics provide the bias currents for the \squid s
and condition all analog signals entering the cryostat by attenuating them and filtering 
them for RF.

\subsubsection{Series Array SQUID} 

Each bolometer module is connected directly to the input coil of a
series-array \squid\ manufactured at NIST~\cite{NIST-arrays}. Each
array consists of 100~individual \squid\ elements with their input
coils, feedback coils, and outputs connected in series. The input
signals add constructively while the output noise voltage of the
individual elements adds incoherently, resulting in a signal-to-noise
ratio that is enhanced by $\sqrt{n}$, where $n=100$ is the number of
elements. The array noise is 3.5 pA/\rtHz\ referred to the input coil
and the bandwidth is 120~MHz.


Each \squid\ element has two inductor coils that were intended for use
separately as input coil and feedback coil. For shunt-feedback, the
input coil is used for both functions. This coil array has an input
inductance $L^{SQ}_{IN} \simeq 150~nH$ and mutual inductance for each
8-turn element $M^{SQ}_{IN} \simeq 80~pH$.

The response function for an open-loop \squid\ may be approximated 
as sinusoidal, with output voltage
\begin{equation} \label{e_squid_response}
  V^{SQ}_{OUT} \simeq V_{PK} \sin ( 2 \pi { I_{IN} } / { I_{\Phi_0} } ) ,
\end{equation}
where $I_{IN}$ is the current through the input coil, $V_{PK}$ is the
peak output voltage of the device, and $I_{\Phi_0}$ is the input
current required to provide a quantum of flux $\Phi_0 = h / 2e$
through the \squid. For our devices $I_{\Phi_0}\approx 25~ \mu$A and
$V_{PK}\approx 1.5-3.5$~mV depending on fabrication reproducibility, operating
temperature, and tuning. One can choose the operating point along the
sinusoidal response curve using a static flux bias. For the shunt-feedback
circuit, the devices are operated on the mid-point of the falling
edge (see \S\ref{s_squid_tuning}). The derivative of
Equation~\ref{e_squid_response} at this point is the transimpedance,
$Z^{SQ}_{trans}\simeq -500 V/A$, in good agreement with the measured
performance of the devices.

\subsubsection{Shunt-feedback Circuit}

Shunt-feedback (Figure~\ref{f_SQUID_FLL} ) is essential for
linearizing the \squid\ response and reducing the input reactance of
the FLL circuit so that the bolometer impedance dominates the bias
loop.
Had the more common feedback
configuration with separate feedback and input coils been used, it
would increase the FLL circuit input impedance substantially,
spoiling the bolometer voltage bias and relegating it to an unstable
regime of electro-thermal feedback.

Shunt-feedback is rarely used for practical \squid\ applications,
because it couples the the strong input coil directly to the room
temperature 
electronics, where pickup from digital components or RF signals are
more difficult to shield against. Implementation of a shunt-feedback
system requires careful filtering and shielding to prevent outside
unwanted signals from reaching the \squid\ input coil through the
feedback line. Since this implementation is uncommon, the basic
equations that characterize the shunt-feedback circuit operation are
presented below.

An open-loop \squid\ has input reactance 
$Z^{SQ}_{IN} \simeq j \omega L^{SQ}_{IN}$ 
(the reflected impedance of the \squid\ is small at these frequencies
and has been omitted), where $\omega$ is the angular frequency.
The forward voltage gain of the open-loop \squid/amplifier cascade is
\begin{eqnarray}
G_{FWD} && = \frac{ V^{AMP}_{OUT} }{ V^{SQ}_{IN} } 
              =  \frac{ - I^{SQ}_{IN} | Z^{SQ}_{trans} G^{AMP} | }
                       { j  I^{SQ}_{IN} \omega L^{SQ}_{IN} } \nonumber\\
            &&  = \frac{ j | Z^{SQ}_{trans} G^{AMP} | } { \omega
              L^{SQ}_{IN} } ,
\end{eqnarray}
where $V^{AMP}_{OUT}$ and $V^{SQ}_{IN}$ are the voltages at the
amplifier output and across the \squid\ input coil respectively,
$I^{SQ}_{IN}$ is the current through the \squid\ input coil, and
$G^{AMP}\simeq 250$ is the non-inverting amplifier forward gain. The \squid\
transimpedance is written $Z^{SQ}_{trans} = - | Z^{SQ}_{trans}|$ to
make explicit its operation in inverting mode. It is a simple
extension to include the phase response of the amplifier.

When the switch in Figure~\ref{f_SQUID_FLL} is closed to complete the
feedback loop, input current is split between the \squid\ coil and the
feedback resistor, the latter of which has its impedance reduced by
the forward gain of the circuit so that the total input impedance
$Z^{tot}_{in}$ is the parallel combination 
\begin{equation}
  Z^{tot}_{in} = Z^{SQ}_{IN} ||
               \frac{ R_{FB} }{ 1 - G_{FWD} } .
\end{equation}
For this system $G^{FWD}\stackrel{>}{\sim} 10^5$ across a 1~MHz bandwidth and a
computer controlled switch allows feedback resistors
$R_{FB}=3.3$k$\Omega$, 5k$\Omega$, or 10k$\Omega$ to be used. The
circuit input impedance is 22-69~m$\Omega$ in the frequency range
300-900~kHz for the $R_{FB}=10$k$\Omega$ feedback setting most
commonly used for SPT and APEX-SZ.
%

The \squid\ response function (Equation~\ref{e_squid_response})
defines its open-loop dynamic range as 
${ I_{IN}^\mathrm{max,\ open\ loop} } = { I_{\Phi_0} }/2$.
The shunt-feedback loop-gain 
\begin{equation}
A^{SQ}_\mathrm{loop} = \frac{ -|Z^{SQ}_{trans}| \times G^{AMP} }{ R_{FB} }
\end{equation}
defines the extent to which the current through the \squid\ coil is
canceled by negative feedback, extending the \squid\ dynamic range to
\begin{equation}\label{e_FLLDynamicRange}
{ I_{IN}^\mathrm{max} } = \frac{ I_{\Phi_0} }{2} \left( 1 -
  \frac{2}{\pi} A^{SQ}_\mathrm{loop} \right)
\end{equation}
and vastly improving its linearity. For the \fMUX\ implementation,
$A^{SQ}_\mathrm{loop} \simeq$ 38, 25, 12.5 for $R_{FB}=$ 3.3k$\Omega$,
5k$\Omega$, and 10k$\Omega$ respectively. This results in a dynamic
range that is extended by a factor 23, 15, and 7, respectively.

When the \squid\ dynamic range is exceeded, the circuit
``flux-jumps'' such that the \squid\ is locked in a state one
 flux quantum or more away from its original locking point. This is similar
to an operational amplifier hitting its rail and `sticking' there. The FLL has
drastically deteriorated dynamic range and linearity in its
flux-jumped state. A flux-jump can be caused by electrical pick-up
(typically of fast, digital signals), electrical glitches, or (very rarely)
by a large magnitude electronic noise fluctuation. 

For the 120 \squid\ arrays in SPT, the
flux-jumping rate is typically between
0 to 1 \squid\ arrays per 24~hour observing cycle. There are
several means of recovering from a flux-jump. The simplest method is
to `reset' the loop by either opening and closing it again (using the
switch located in the feedback path, Figure~\ref{f_SQUID_FLL}) or removing and
reinstating the \squid\ bias. These `reset' methods temporarily set
the loop-gain $A^{SQ}_\mathrm{loop}$ to zero. This `reset' only works if the
input current through the \squid\ coil is zero, otherwise the
\squid-loop will not be locked at the original locking point. A
complication for this method is that, with $A^{SQ}_\mathrm{loop} =0$,
the input impedance $Z_{in}^{tot}$ of the FLL is altered, which alters
the resonant frequencies of the $LC\Rbolo$ resonant circuits. This can
spoil ETF, causing the bolometer to oscillate and latch in a
superconducting state. For the \fMUX\ system, the \squid\ input is
always non-zero, due to sky-signal and imperfections in setting the
nulling amplitude. Consequently, instead of `resetting' the FLL,
\squid-jumps are 
recovered from by injecting a corrective current through the \squid\
flux bias wire, and increasing this current until the \squid\ jumps
back to the original locking point.

\subsubsection{Flux Locked Loop Stability}

To maintain stability in the \squid\ FLL circuit, phase shifts along
the feedback loop must provide for negative feedback up to the
frequency where the loop-gain bandwidth product falls below unity,
38~MHz for $R_{FB}=3.3$k$\Omega$. The room temperature operational
amplifier, with a   
bandwidth of 1~MHz, induces a 90$^\circ$ phase shift, allowing for an
additional margin of 45$^\circ$. This places stringent requirements on
the length of the wires connecting the \squid s to the room
temperature amplifier since a 0.66~m round-trip propagation delay amounts
to 45$^\circ$ at 38~MHz. A one-way wiring length of 0.2~m is used for
APEX-SZ and SPT. In the absence of phase shifts induced by strays,
this lead length would allow stability. In practice, phase shifts
arise from other non-idealities, such as the parallel resonance formed
by the capacitive coupling of the MUX inductor to its flux-focusing
washer or the wiring strays discussed in \S\ref{s_cold_strays}.

To improve the FLL stability, a lead-lag filter is introduced
and connects the output of the \squid\ through a capacitor and
resistor to ground, see Figure~\ref{f_SQUID_FLL}.
At
high frequency, this $RC$ shunt-filter looks purely resistive and acts
as a voltage divider with the \squid\ output, attenuating the
loop-gain. The filter introduces a phase shift at intermediate
frequency, when the reactance of the capacitor is similar to the
filter resistance. The filter capacitance is 1~nF and its resistance
is set by a 0.2~m Manganin wire, about 15~$\Omega$. This filter
introduces a frequency dependence on the \squid\ transimpedance,
attenuating it by about 30\% at the high end (900~kHz) of the \fMUX\
carrier bandwidth. It also introduces a phase shift inside the \fMUX\
carrier bandwidth. These two effects contribute to a rise in the noise
from the room temperature electronics at high bias carrier frequency,
discussed in \S\ref{s_noise}.

\subsubsection{SQUID Electronics Implementation}

The \squid\ electronics are implemented on two circuit boards: (1) a
\squid\ mounting board that houses the \squid s at 4~K and provides
magnetic shielding, and (2) a room temperature \squid\ Controller that
houses the FLL operational amplifiers and \squid\ bias current
generation circuits, and conditions signals as they enter or exit the
cryostat.

\begin{figure}
\includegraphics[width=0.45\textwidth,clip=]{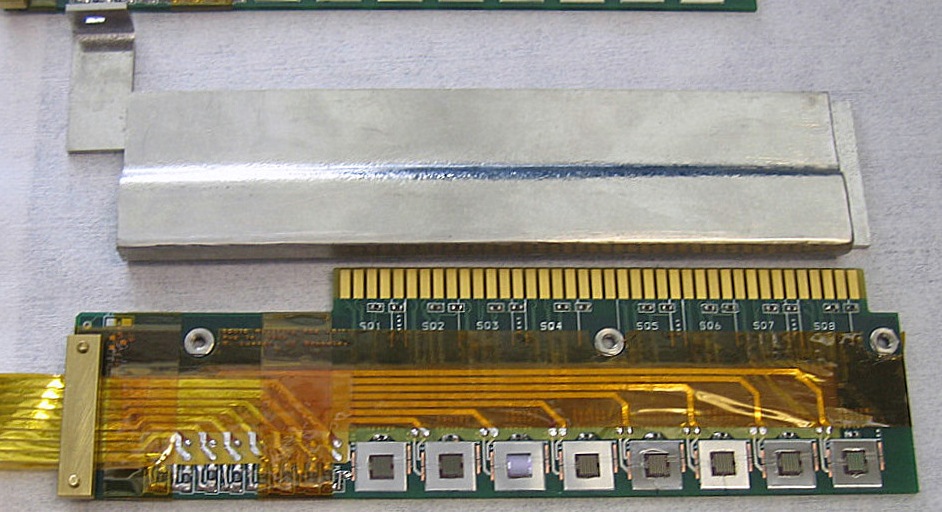}
\caption{ \label{f_squidboard_photo} (Color online) Photo showing the \squid\
  mounting board, housing eight NIST \squid\ series arrays and
  the bias resistors for the bolometers.
  The \squid s are mounted on top of a Niobium film, visible in the photo.
  A magnetic shield enclosure, fabricated from the high magnetic
  permeability material Cryoperm, is shown above the
  circuit board.}
\end{figure}

The \squid\ mounting boards, shown in 
Figure~\ref{f_squidboard_photo}, are 0.15$\times$0.05~m$^2$ 
custom printed circuit boards housing eight 
\squid s each. \squid s are sensitive to changes in magnetic field, so
that effective magnetic shielding is essential. Since the system employs
series array \squid s, it is important to attenuate spatial variations
as well as time variations in the magnetic field. This is because a
spatial variation across the array would spoil the coherence of the
individual \squid\ elements.

Each \squid\ is mounted with rubber cement on top of a
10$\times$10~mm$^2$ 
niobium film that is epoxied to the \squid\ mounting board. Niobium is
a Type~II superconductor and serves to pin magnetic field lines,
greatly attenuating their time variability. The entire circuit board
is placed inside an enclosure built from a
material that has a high magnetic permeability at cryogenic
temperatures (Cryoperm, Vacuumschmelze GmbH, 
Gr\"uner Weg 37 D-63450, Hanau, Germany).
This attenuates the amplitude of the field lines,
reducing both the temporal and spatial variability. The enclosure is
open on two sides through narrow ``chimneys'' to allow wiring
connections to the room temperature electronics through an edge
connector, and to the cold multiplexer components through a broadside
coupled stripline that is soldered directly to the board.

The 30\,m$\Omega$ bolometer bias resistors ($R_{bias}$ in
Figure~\ref{f_fmux_implementation}) are located on the \squid\
mounting boards. This is a compromise, because the 4~K temperature
means the bias resistors contribute about 2.5~pA/\rtHz\ of noise
referred to the \squid\ input. This would be reduced to a negligible
level had this bias resistor been placed at 0.25~K
with the sub-kelvin multiplexer components. By keeping the resistor at
4~K, we are able to connect to each multiplexer comb with just two
wires instead of four, reducing the heat load on the sub-kelvin stage.

The \squid\ controllers are 180$\times$170~mm$^2$ custom printed
circuit boards. One board provides the control circuitry for eight
\squid s.  Each \squid\ control circuit includes a high gain-bandwidth
operational amplifier with low (1~nV/\rtHz) input noise and two
feedback resistors (5k$\Omega$ and 10k$\Omega$, $R_{FB}$ in
Figure~\ref{f_SQUID_FLL}) that can be enabled individually or in
parallel via software with an analog CMOS switch (adjacent to $R_{FB}$
in Figure~\ref{f_SQUID_FLL}) that is selected for its low charge
injection. A DAC provides four software programmable voltages for each
\squid\ that are used to (1) provide a current bias, about 150~$\mu$A,
through the \squid, (2) provide a flux bias current, 0-25~$\mu$A, to
the \squid\ input coil that is used to set the magnetic field through
the \squid, (3) zero the offset of the room temperature op-amp, and
(4) provide a current to a heater resistor next to the \squid\ that
can be used to heat the \squid\ above its superconducting transition.

Analog bolometer bias carrier signals and nulling signals are received
from the \oscdemod\ boards through isolation transformers on
differential twisted pair cables, and attenuated on the \squid\
controller board before being sent into the cryostat. The FLL output
signals are transmitted from the \squid\ controller boards to the
\oscdemod\ boards using differential drivers feeding twisted pair
cables.

\squid s are very sensitive to digital pickup, so the digital
circuitry on the \squid\ controller boards is isolated from the analog
circuitry by separating the local grounds and coupling the connections
from the digital to analog circuitry through resistor-capacitor
filters. Balanced connections between the SQUID controller and the
demodulator circuit boards also suppress digital interference from the
crate. Overall, digital noise pickup is negligible.

\subsubsection{\squid\ Setup and Tuning}\label{s_squid_tuning}

The sorption fridge that cools the sub-kelvin stage of the system
has a hold time of about 36 hours. Each time this fridge is cycled,
the \squid s are tuned to configure them at their optimum operating
point.

As the mainplate is cooled to 4~K, the \squid s cross their
superconducting transition relatively slowly, with a thermal gradient
across the series arrays. This gradient can transform time varying
magnetic fields into spatially varying trapped flux across the series
array. \mynewtext{The first step in the tuning process is to raise the
  temperature of each \squid\ by providing current to a heater
  resistor located next to it, bringing the device into the normal
  state and then allowing it to cool quickly, reducing the probability
  that time varying fields will be converted into spatially trapped flux.}

\begin{figure} \includegraphics[width=0.5\textwidth,clip=]{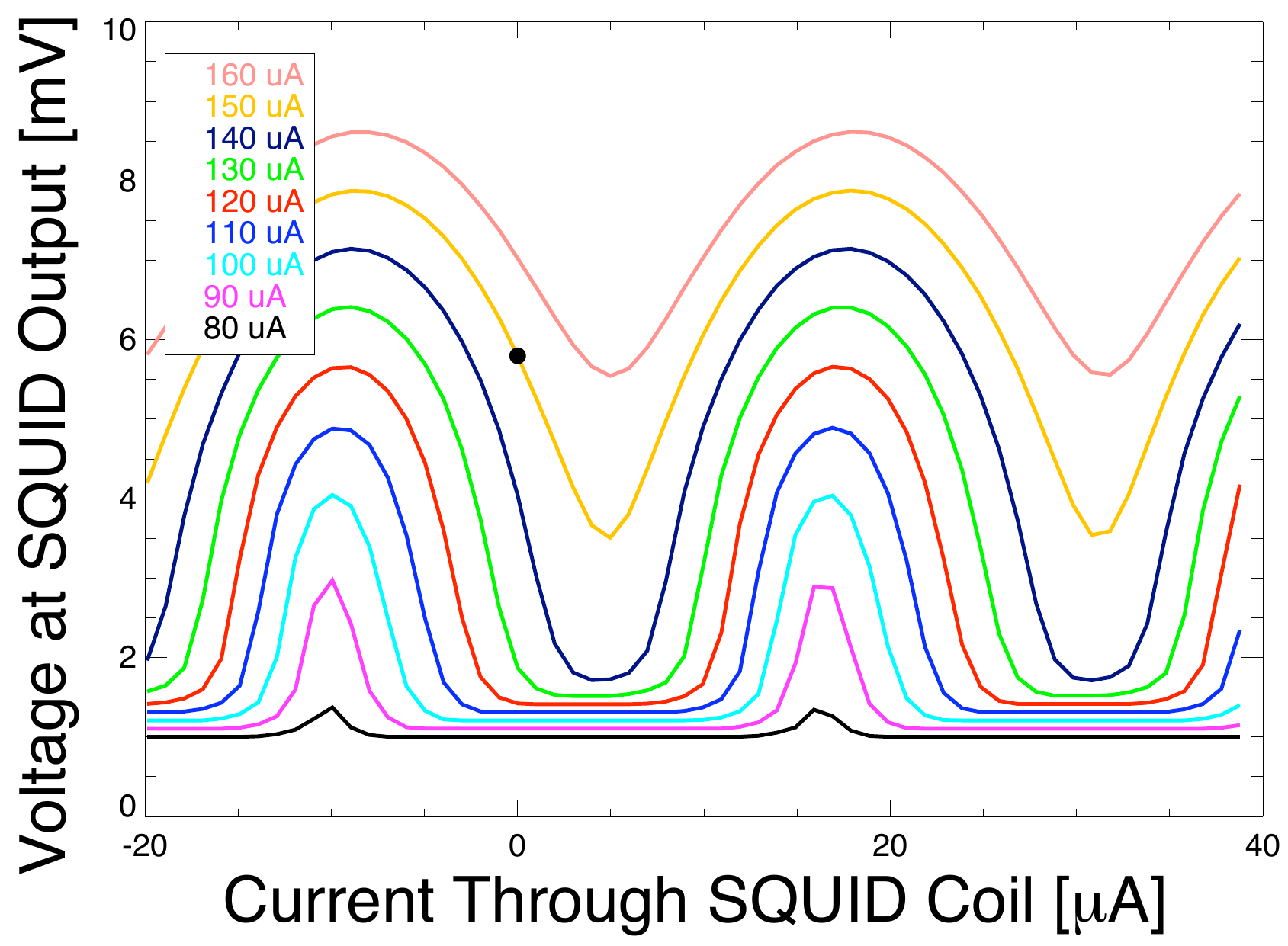}
  \caption{ \label{f_squidVphi} (Color online) The \squid\ voltage response to a 
    linearly increasing current through the \squid\ input coil
    (called a ``V-phi'' curve)  is
    shown for several choices of the \squid\ bias current ranging from
    80~$\mu$A (bottom curve) to 160~$\mu$A (top curve). 
    The chosen \squid\ operating point is marked by a closed circle.
}
\end{figure}

The next step is to map out the \squid\ output voltage response to a
current through its input coil (approximated by
Equation~\ref{e_squid_response}) as a function of the \squid\ bias
current, as shown in Figure~\ref{f_squidVphi}. The bias current choice
defines which of these ``V-phi'' curves the \squid\ operates on.

The derivative of the V-phi curve is the \squid\ transimpedance
$Z^{SQ}_{trans}$. It defines the \squid\ small signal response and is
the quantity that refers the room temperature electronics noise back
to an equivalent noise current through the bolometer. Ideally,
$Z^{SQ}_{trans}$ would be maximized.  The peak-to-peak amplitude
relates to the \squid's large signal response. The distance between
peaks in each V-phi curve is the current required to produce a flux
quantum through the \squid\ coil, and is independent of the bias
current. For stable \squid\ operation, a smooth curve with a high
level of symmetry around the chosen flux bias point is preferred. A
typical choice for the \squid\ operating point is indicated by the
small red circle 
on the 150~$\mu$A \squid\
bias curve (2nd from top). This represents a trade off between the
criteria listed above.

Once the \squid\ response has been mapped and the bias currents
chosen, the feedback loop is closed so that the FLL is in its low
input impedance state, allowing for bolometer operation.

\subsection{Demodulator and Digitization}
\label{s_demodulator}

\begin{figure}
\includegraphics[width=0.5\textwidth,clip=]{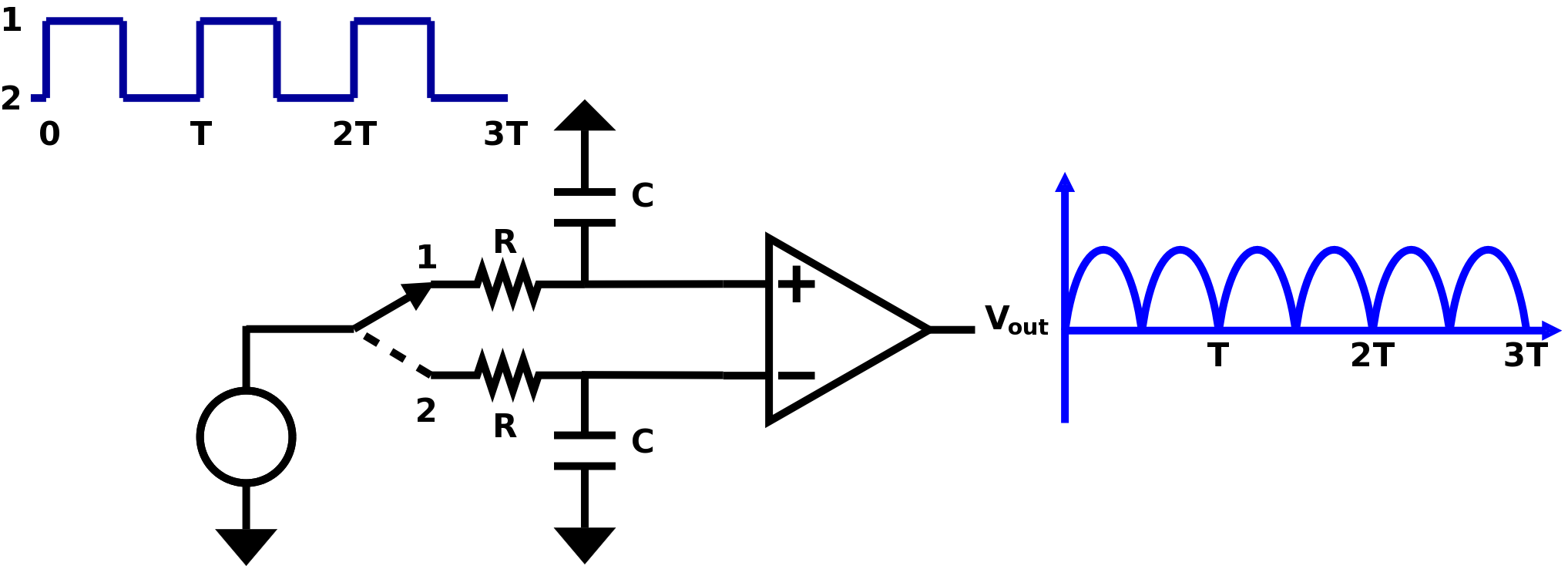}
\caption{ \label{f_demodulator} 
Block diagram of the demodulator circuit.
}
\end{figure}

Each bolometer channel has its own analog demodulator that locks into
the bias carrier and mixes the sky-signal down to base-band. This
demodulator is referenced with the same analog sine wave that biases
its bolometer, serving to cancel the clock-jitter sideband
noise. Sixteen demodulator channels are housed on each \oscdemod\
board, as described in \S\ref{s_carrier_gen}.

Eight demodulators (half of an \oscdemod\ board) monitor each
multiplexer module, consisting of seven bolometer channels. Seven of
the demodulators are phase-locked to the in-phase $I$ component of the
seven bolometer carriers. The system was originally conceived with 8
bolometers per module, but this has been reduced to seven so that the
eighth ``helper'' demodulator can be used periodically to measure the
$Q$ component (90 degrees out of phase) of each channel. Being able to
simultaneously observe both $I$ and $Q$ components facilitates fast setup
of the nulling signal amplitude and phase discussed in \S\ref{s_bolo_tuning}.

Each module of eight demodulators is AC coupled to the differential
\squid\ controller output through capacitors. An input transformer
suppresses common-mode signals. In front of the coupling capacitors, a
separate analog to digital converter (ADC) samples the \squid\ static
offset at 1~kHz, which is necessary for tuning the \squid s and
monitoring for flux-jumps.

The signal from an individual bolometer is mixed down to base-band by
multiplying it with a unit amplitude square wave. This is achieved
(Figure~\ref {f_demodulator}) by passing the reference carrier
sinusoid through a comparator and using its output to control an
analog switch that acts as a synchronous rectifier on the input
waveform. The output of this rectifier is sent to a pair of low-pass
$RC$ integrator circuits that remove the high frequency out-of-band
signals produced by the other bolometer channels. The rectifier operates in a
doubly-balanced mode, with differential input signals, not shown in
Figure~\ref{f_demodulator}, as well as the differential output. This
implementation is built up from discrete components and has excellent
distortion performance because there are no non-linear elements. An
instrumentation amplifier, with gain that is software selectable
over two orders of magnitude, takes the difference between the two
integrators. The preceding stages have sufficient gain to override the
instrumentation amplifier's low-frequency noise.
The amplified base-band signal is anti-alias filtered
with an 8-pole active low-pass filter operating at 400~Hz and
digitized with a 12-bit ADC at 1~kHz.

A Field Programmable Gate Array (FPGA) on each \oscdemod\ board
assembles the outputs from the ADCs and sends the data via an 8-bit
parallel low voltage differential signal (LVDS) connection to the
readout control computer. A software low-pass filter with a 35~Hz
pass-band is applied to the data, which is then down-sampled to 100~Hz and
written to disk.

\subsection{Readout Electronics Crates, Interface, and Power Consumption}
\label{s_cabling_enclosure_power}

\begin{figure}
\includegraphics[width=0.5\textwidth,clip=]{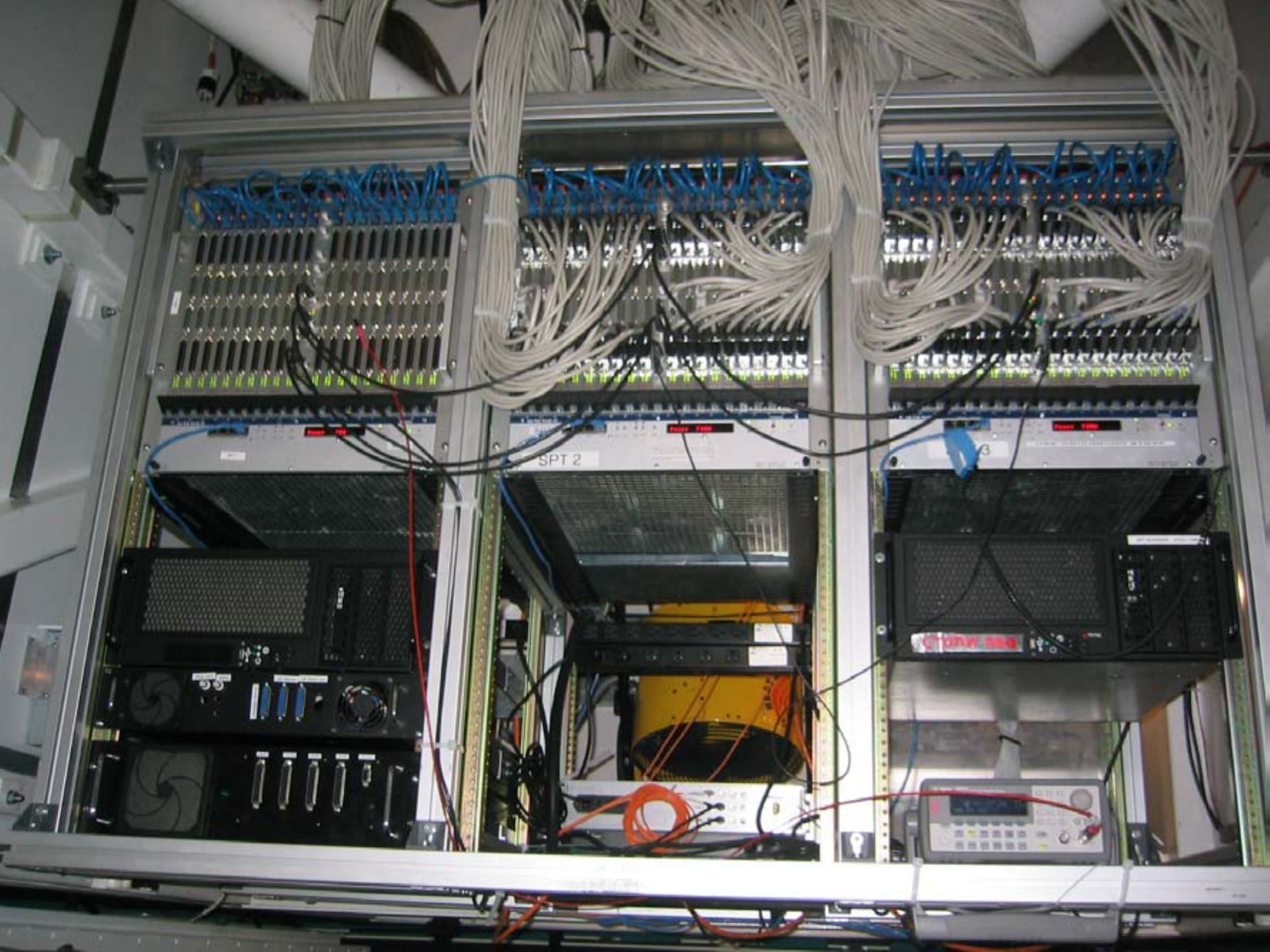}
\caption{ \label{f_spt_readout_photo} (Color online)
Photo showing the three readout electronics crates (top half of each
rack) housing the oscillator/demodulator circuit boards in the cabin
of the South Pole Telescope. Control computers, cryogenic
housekeeping, timing, and network electronics are also visible in the
bottom half of the racks.}
\end{figure}

The 16-channel \oscdemod\ circuit boards
(Figure~\ref{f_oscdemod_photo}) are $340\times 360$~mm$^2$ 
in size and 20~circuit boards are packaged together in a 9U
Versa Module Eurocard (VME)
crate with custom backplanes that provide power
to the boards 
and connect to the control computer. Each crate is convectively cooled
with fans and has an integrated power supply. The power consumption is
about 900~W per crate including the power delivered to the \squid\
Controller electronics described above. Three crates (2700~W) house
the oscillator/demodulator electronics for the 840~bolometer channels
of SPT (photo Figure~\ref{f_spt_readout_photo}). The 280~channel
APEX-SZ system requires one crate (900~W).

The center slot of each crate is used for a clock distribution
circuit board that sends a central 40~MHz system clock and a
once per second GPS-derived timestamp across the backplane to each
\oscdemod\ board. The system
clock and timestamp signals can be daisy-chained across several clock
distribution boards in different crates.

The system is interfaced to a control computer, running the Scientific
Linux operating system. Command signals are received from the control
computer through a single RS485 serial signal distributed across the
backplane and encoded using the communications
protocol \partno{Modbus}{see http://www.modbus.org}. Bolometer data,
sampled at 1~kHz, flow from the readout crate backplanes to the
control computer through an 8-bit parallel low voltage differential
signal (LVDS) digital output bus operating at 40 MHz. 

A digital token is passed sequentially from board to board through a
dedicated hardware port to define which circuit board has control of
the output bus. This protocol allows all boards in the system to share
the same digital output.

\subsection{Bolometer Tuning} \label{s_bolo_tuning}

The bolometers are tuned by adjusting the
bias voltage for each detector and measuring changes in current
through the bolometer using the \oscdemod\ boards. The tuning algorithms are implemented in software running on the readout control computer.

The bolometer tuning takes place after the \squid s have been tuned
(see \S\ref{s_squid_tuning}) and the $LC\Rbolo$ resonances have
been determined with a network analysis (see
\S\ref{s_cold_mux_implementation}). 

The first step is to cool the bolometer stage to about $\sim$800\,mK,
which is above the TES transition temperature but below that of the
aluminum leads connecting the detectors to the cold multiplexer.

Once this temperature is reached, the detectors are provided with a
sufficiently large electrical bias that the devices will remain above
their transition temperature when the bolometer stage is cooled to its
operating temperature of 250\,mK. Since the \squid\ dynamic range is
not sufficient to handle seven bias carriers simultaneously, the seven
detectors in a multiplexer module are biased sequentially. After each
bias is turned on, a nulling sinusoid is adjusted to cancel the
carrier signal at the \squid\ input, before the next TES bias is
activated.

The optimum nulling sinusoid amplitude and phase are determined using
measurements of both the $I$ and $Q$ components of the demodulated
bolometer current as provided by each bolometer's demodulator and the
one-per-multiplexer module ``helper'' demodulator. After the optimum
phase is determined, the ``helper'' demodulator is no longer needed,
and can later be used to tune another channel. Once all seven channels
in the multiplexer module are biased and nulled, the bolometer stage
temperature is lowered to its operating temperature of 
$\sim$250\,mK. The
nulling amplitudes are fine-tuned at this temperature to correct for
slight impedance changes in the circuit.

\begin{figure} \includegraphics[width=0.5\textwidth,clip=]{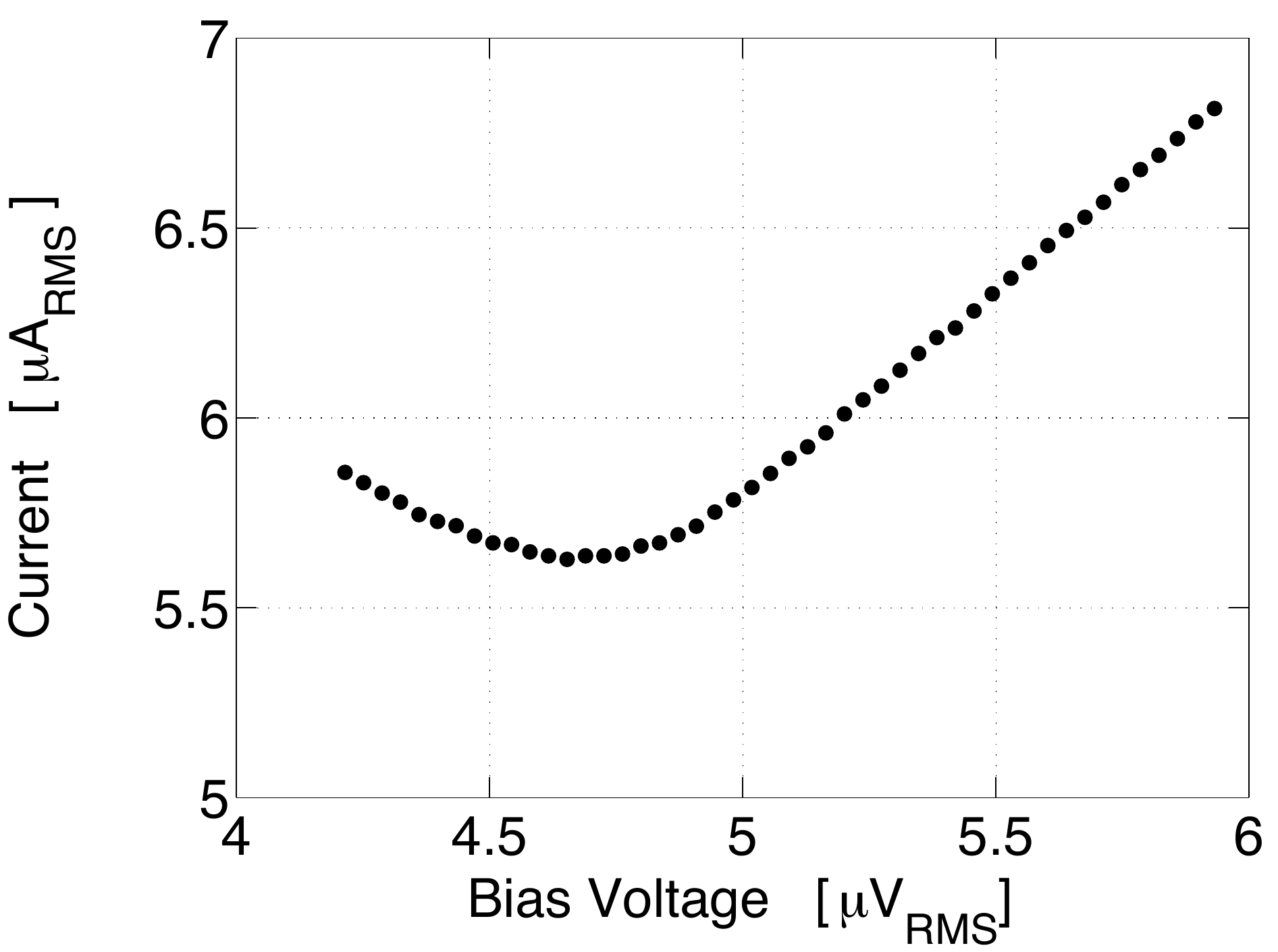}
  \caption{ \label{f_boloiv} A current vs. voltage curve is
    shown for an SPT TES bolometer operating dark at a bath 
    temperature of
    0.29~K. In the upper-right portion of the curve, the TES
    electrical bias power is sufficient to hold the device 
    above its superconducting transition, with constant resistance 
    $\Delta V/\Delta I$.  
    As the bias voltage is
    lowered, the device falls into its transition and is held there by
    electro-thermal feedback (ETF).  ``Turn-around'' is defined as the
    minimum in the curve, about 4.7 $\mu$V, when the dynamic
    resistance $\partial V/\partial I$ transitions from positive to
    negative. 
}
\end{figure}

The seven bolometers in each module are lowered into their transition
sequentially by decreasing the electrical bias power. For a given
bolometer, the bias voltage amplitude is decreased while iteratively
adjusting the nuller amplitude and phase to keep the current through
the \squid\ zero (this corrects for the voltage drop across the
\squid\ input, p5 in Figure~\ref{f_mux_strays}, as discussed in
\S~\ref{s_cold_strays}). This traces out the bolometer current vs.\
voltage curve (Figure~\ref{f_boloiv}). In the upper-right portion of
the curve, the TES electrical bias power is sufficient to hold the
device above its superconducting transition. As the bias voltage is
lowered, the device falls into its transition and is held there by
ETF. ``Turn-around'' is defined as the minimum in the curve (4.7
$\mu$V in Figure~\ref{f_boloiv}) when the dynamic impedance $\partial
V/\partial I$ transitions from positive to negative (the average TES
resistance $\Delta V/\Delta I$ is always positive). At this point the
detector ETF loop-gain \loopgain\ is unity. Once the TES reaches its
target operating resistance, the bias voltage is held constant and the
current is once again nulled at the \squid\ input. All seven
bolometers in the multiplexer comb are tuned sequentially in this
manner. The tuning of bolometers in different multiplexer modules
across the system is performed in parallel. For astronomical
observations with SPT (APEX-SZ), the TES sensors are typically
operated at 60-80\% (80-90\%) of their normal resistance corresponding
to typical ETF loop-gains \loopgain$\sim$15-20.

After the electrical bias power of each detector in the system has
been decreased so the detectors are operating in their superconducting
transition, the demodulator gain is increased from its high dynamic
range setting to one of its low noise settings, that has
33-100 times larger gain. The nulling amplitudes are fine-tuned and astronomical
observations can commence.

When detector loading conditions change, for example, when a large
change in observing elevation is made, adjustments in the nulling may
be necessary. Significant changes in detector loading may necessitate
re-tuning of the detectors.


\section{System Performance} \label{s_performance}

In this section, measurements of the system performance---including
noise, sensitivity, and cross-talk---are presented.

\subsection{System Setup}\label{s_system_setup}


For SPT and APEX-SZ, the bolometer tuning (described in more detail in
\S~\ref{s_bolo_tuning}) is integrated with the cycling of the
sub-kelvin sorption fridge to minimize the total system setup time.
The sorption fridge is cycled every 24-36\,hours, elevating the
temperature of the bolometers and \squid s to above their
superconducting transition. This necessitates a re-tuning of the bias
parameters for both. The process used for SPT is outlined below.

The \squid s are returned to their operating temperature mid-way
through the cryogenic cycle and are tuned (\S~\ref{s_squid_tuning}) in
parallel with the second half of the cycle. At this point, the
bolometers are above their superconducting transition. The \squid s
are first heated for 30 seconds to release trapped flux and are
subsequently allowed to cool quickly through their transition.
Following this cooling, it takes about 25 minutes for the \squid s to
return to their base operating temperature. Next, voltage-current
curves (Figure~\ref{f_squidVphi}) are measured for each \squid\ to
determine their optimal flux and current bias. For SPT, tuning all 120
\squid s takes $\sim$25\,minutes, and is performed in parallel with
the sorption fridge's cycle, therefore coming at no cost to the
observing efficiency.

Towards the end of the cryogenic cycle, the bolometer stage is left at
$\sim$800\,mK so that the detectors are above their superconducting
transition. The control computer commands the readout system to bias
the detectors with enough electrical power to hold the TES normal
and tune the nulling current to cancel the bias at the \squid\
input (\S \ref{s_bolo_tuning}); this takes about 10\,minutes. After
this, it takes about 2~hours for the sorption fridge to cool the
bolometer stage to its $\sim$250\,mK base temperature. Once the
cooling is complete, it takes about 15~minutes to adjust each
detector voltage bias to the desired point in its superconducting
transition, and then to re-adjust the nulling current (\S
\ref{s_bolo_tuning}). An I-V curve is recorded for each detector
(e.g.,\ Figure \ref{f_boloiv}). The total time elapsed for
readout/bolometer system tuning and setup is about 25~minutes, not
counting the time (several hours) that would otherwise be spent
cycling the cryogenics.

SPT has 724~operational detectors. Others have wiring or other
defects that prevent their operation. Typically 650$\pm$15 of these
detectors show good performance each tuning and are used for
astronomical observations. Detailed statistics and characterization of
end-to-end receiver performance are presented in a separate
publication~\cite{BensonSptReceiver}.

\subsection{Electrical Cross-Talk}

Channel-to-channel cross-talk can occur through optical signals or
electrical signals. Optical cross-talk between adjacent bolometers in
SPT and APEX-SZ is typically about 1\% and is determined by the
optical design of the experiment and the arrangement of detectors
within a silicon wafer. The goal is for electrical cross-talk arising
in the readout system and discussed in \S~\ref{s_cross_talk} to be small
in comparison to this optical cross-talk.

Electrical cross-talk in the multiplexer readout has been measured
both in the laboratory and during sky-observations. \mynewtext{ For
  the laboratory measurement~\cite{2005ApPhL..86k2511L}, one detector,
  mounted in its own light-tight enclosure, was excited with light
  from an LED. Five other sensors were read out on the same \fMUX\
  module, but mounted in a separate light-tight enclosure.}
The absence of any measurable signal in the five dark channels
established an upper limit of 0.4\% electrical cross-talk.

Cross-talk has been characterized with on-sky observations of RCW38,
MAT5A, Venus and Jupiter with the South Pole Telescope.
Though we cannot distinguish
between positive cross-talk that arises from optical versus electrical
sources, negative cross-talk should be dominated by electrical
signals. The signal-to-noise levels for RCW38 and MAT5A are
substantially lower than for the planets, and no negative cross-talk
is observed. Using the observations of Venus and Jupiter, which produce
signals that are large enough for small cross-talk signals to be
observed in neighbors, a median cross-talk
between neighbors of 0.3\% is observed. This is consistent with the
expectation of 0.25\% and 0.3\% from mechanisms (2) and (3) described
in \S~\ref{s_cross_talk}.

\subsection{Noise Performance}\label{s_noise}

In this section, the measured noise levels for a dark detector in the
laboratory and an SPT detector observing the sky are compared with the
theoretical expectations.


The noise contributions are divided into three categories: photon
noise, bolometer noise, and readout system noise. Within each
category, a contribution may arise as a current or power source. Sky
signals arise as power sources (photons converted to heat), so that
noise 
equivalent power (NEP) is the relevant metric. Other sources, such as
the \squid\ noise, are constant in terms of an input referred current,
and are labeled here as current noise sources. The NEP of
current noise contributors depends on the device parameters and
operating conditions and is converted from current to NEP by
multiplying the \squid-input referred current noise by the inverse of
the detector responsivity, $\sim -V^\mathrm{RMS}_\mathrm{bias}/\sqrt{2}$ (see
Equation~\ref{e_av_responsivity} in Appendix~\ref{s_demodulator_response})
which is valid in the limit of high loop-gain.

Achieving optimal noise performance with TES bolometers for
astronomical observations requires a careful optimization of detector
and readout system parameters. Most important amongst these is the
power required to saturate the detectors, $P_\mathrm{sat}$. If this power is
many times higher than the absorbed radiation power (or the detector
is dark, so that there is no radiation power), thermal phonon noise
will be high and relatively large voltage biases must be used to
provide the Joule heating necessary to keep the detector in its
transition. The high voltage bias results in an amplification of
current noise sources when they are referred to NEP, so that the
readout component of the noise is greatly enhanced. On the other hand,
if $P_\mathrm{sat}$ is too small, the detector will not have enough
dynamic 
range to operate through a broad range of observing conditions.
Typically $P_\mathrm{sat}$ is set to be roughly twice the expected
absorbed 
incident radiation ($\approx 2 \times \Prad$) .

Before describing the noise contributions below, we note that a mixer
responds differently to noise that is encoded as a modulation on the
carrier and noise that is superimposed on the carrier at the same
frequency (see Appendix~\ref{s_demodulator_response}). This
implementation of the \fMUX\ system uses a square-wave mixer
(\S~\ref{s_demodulator}), which introduces a further distinction
between (a) broadband noise sources that are superimposed on both the
carrier frequency and its harmonics, and (b) narrow-band noise sources
that are superimposed on the carrier but do not extend out to the
carrier harmonics. \squid\ noise is an example of (a) because its
bandwidth is not limited. Bolometer Johnson noise is an example of (b)
because its bandwidth is limited by the $LC\Rbolo$ filter. For a
square mixer, these noise sources are enhanced in the demodulation
process by factors of $\frac{\pi}{2}$ and $\sqrt{2}$ respectively, in
comparison to noise that modulates the carrier. Fortunately, in an
alternating-voltage biased system, power terms receive an enhancement
by a factor $\sqrt{2}$ (with respect to a constant-voltage biased
system) during the modulation process, so that the signal-to-noise 
ratio is nearly unaffected by these factors in comparison to a
constant-voltage biased system. 
Appendix~\ref{s_demodulator_response} presents a detailed discussion
and derivation of the relevant factors.

\subsubsection{Dark Noise}

\begin{figure}
\includegraphics[width=0.475\textwidth,clip=]{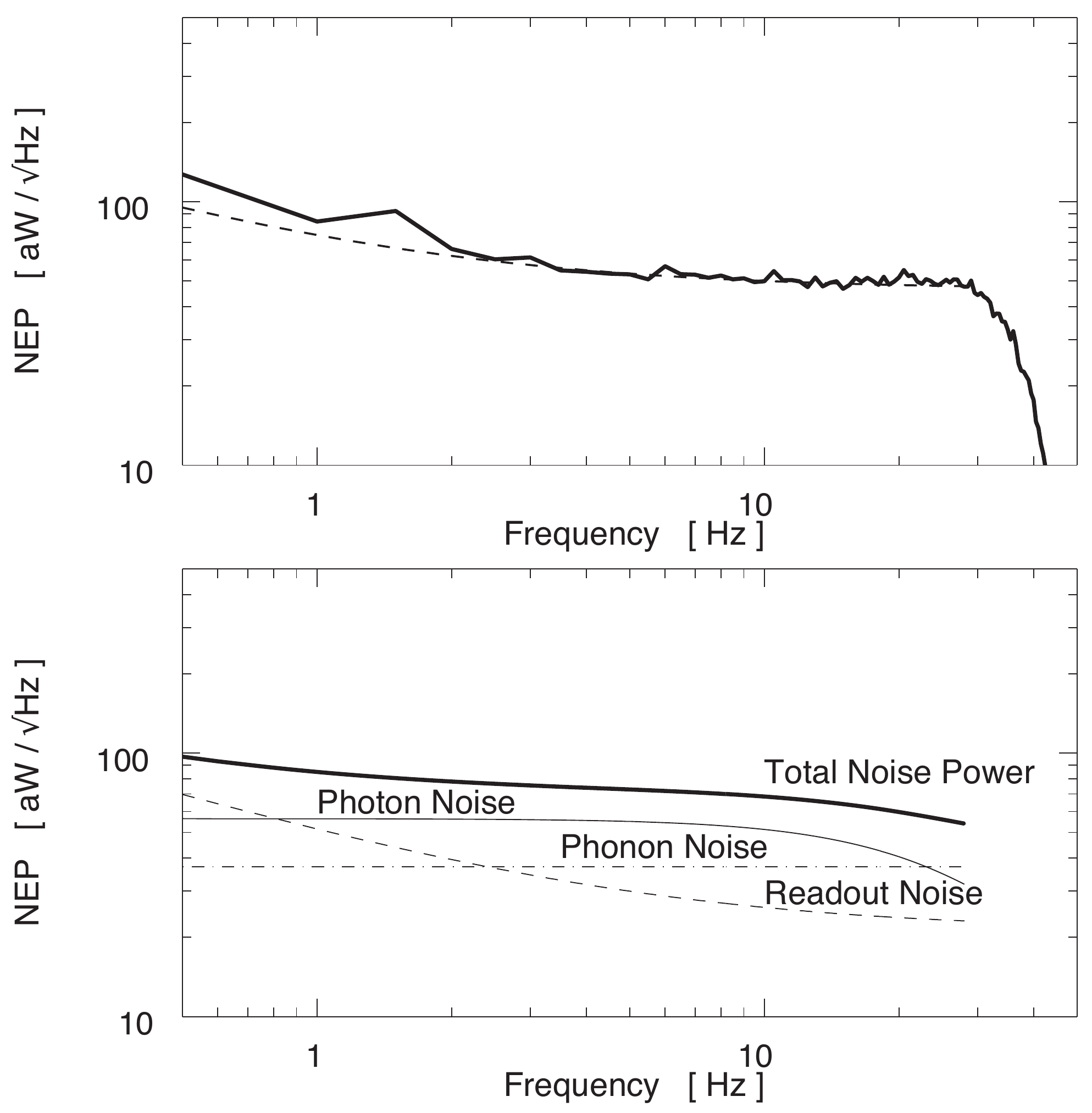}
\caption{ \label{f_darkNoise} 
  The measured noise equivalent power (NEP) spectrum density for an SPT 
  TES bolometer 
  operated at 0.73$R_\mathrm{Normal}$ in the laboratory
  with no incident radiation (``dark'')
  is shown (top, solid line) with the theoretical expectation
  from Table~\ref{t_darkNoise} superimposed (top, dashed line). 
  This configuration represents the worst possible 
  configuration for the readout noise contribution, 
  since the incident power is zero. This
  means the voltage bias, which refers current noise sources to noise
  equivalent power, is maximal. In
  the lower panel the on-sky noise for this detector is estimated
  assuming a total radiation loading of 11~pW and the detector
  parameters of Table~\ref{t_darkNoise}. With radiation power,  the
  detector voltage bias is lower and the readout noise is
  reduced when referred to NEP. }
\end{figure}

In Figure~\ref{f_darkNoise} (top, solid line) the noise equivalent
power (NEP) density is shown for a representative SPT detector
operated in the laboratory with no incident radiation power
(``dark''). The detector has a saturation power of $\bar{G}\Delta T
\simeq 25$\,pW. Its normal resistance 
$R_\mathrm{Normal}=0.98\,\Omega$
and it has been lowered into the transition to
0.73\,$R_\mathrm{Normal}$ with a voltage bias of
4.2\,$\mu$V$_\mathrm{RMS}$ (the current-voltage curve for this
detector was shown in Figure~\ref{f_boloiv}). The roll-off in the
measured noise at 35~Hz is due to the readout system's software low-pass
filter described in \S~\ref{s_demodulator}.

Noise expectations for
this detector are summarized in Table~\ref{t_darkNoise}. Columns are
included for dark operation (zero radiation power, shown in the top
panel of Figure~\ref{f_darkNoise} as a dashed line) and for a
radiation loading of 11\,pW (shown in the bottom panel of
Figure~\ref{f_darkNoise} and discussed in \S~\ref{s_on_sky_noise}),
which is typical of the 150\,GHz detectors for SPT.
Table~\ref{t_darkNoise} also includes the equations and parameters used to
estimate each noise component. A similar table appears in
Ref.~\cite{apexSZ_instrument} for APEX-SZ detectors. The noise
contributions for ``dark'' operation are described below.

\begin{table*}
\setlength{\extrarowheight}{3pt}
\begin{tabular}{lccc} \hline
\em{Detector Parameter:} \\
\hspace{0.5cm} 
TES Transition Temperature & $T_\mathrm{bolo}$=550\,mK   & & \\
\hspace{0.5cm} 
Thermal Conductivity & $\bar{G}=\frac{\Delta P}{\Delta T}\simeq 96$\,pW/K, 
                                   & $g_T=\frac{\partial P}{\partial T} \simeq 165$\,pW/K & \\
\hspace{0.5cm} 
Center Frequency \& Bandwidth & $\nu=$153\,GHz & $\Delta\nu=$38\,GHz & \\ \hline
\em{Detector Setup:} & & \underline{Dark} & \underline{On-Sky} \\
\hspace{0.5cm} 
Bath Temperature & & $T_\mathrm{bath}$=290\,mK &  $T_\mathrm{bath}$=260\,mK \\
\hspace{0.5cm} 
Incident Radiation Power & & $\Prad=0$\,pW 
                                        & $\Prad=11$\,pW \\ 
\hspace{0.5cm} 
TES Normal and operating Resistance 
                                   & $R_\mathrm{Normal}=0.98\,\Omega$ 
                                             & $\Rbolo=0.73$\,$\Omega$ 
                                             & $\Rbolo=0.73$\,$\Omega$ \\
\hspace{0.5cm} 
Bias Voltage    & & $V_\mathrm{bias}$=4.2\,$\mu$V$_{RMS}$ 
                           & $V_\mathrm{bias}$=3.2\,$\mu$V$_{RMS}$ \\
\hspace{0.5cm} 
Loop-Gain        & & \loopgain$\simeq$ 25 & \loopgain$\simeq$ 15 \\
\hline \hline
  Noise Source &   Equation & \multicolumn{2}{c}{NEP (aW/\rtHz)} \\ 
\hline 
                                         & & \underline{Dark} & \underline{On-Sky} \\
\em{Photons:} \\ 
\hspace{0.5cm}
  Shot noise                                 &   $\sqrt{2 h \nu \Prad }$ 
                                                    & 0 & 47 \\ 
\hspace{0.5cm}
  Correlation noise &  $\sqrt{\xi \frac{\Prad^2}{\Delta\nu}}$ 
                              & 0  & $\sqrt{\xi}\cdot$ 56 \\ 
\\
\em{Bolometers:} \\ 
\hspace{0.5cm}
  Thermal Carrier noise ($\gamma_\mathrm{NE}=\valGammaNE$)
         & $\sqrt{\gamma_\mathrm{NE} 4 k_\mathrm{B} T_\mathrm{bolo}^2 g_T}$
         & 37 & 37 \\
\hspace{0.5cm}
            Johnson noise  &    
             $\sqrt{4 k_\mathrm{B} T_\mathrm{bolo} \Pelec}/\loopgain$        &
         28/\loopgain & 21/\loopgain \\ 
 \\ 
\em{Readout:} \\ 
\hspace{0.5cm}
  Warm electronics &$\frac{\pi}{2}\cdot$ 3.7\,\pArtHz $\frac{V_\mathrm{bias}}{\sqrt{2}}$ & 17 & 13 \\ 
\hspace{0.5cm}
                   SQUID &$\frac{\pi}{2}\cdot$ 3.5\,\pArtHz $\frac{V_\mathrm{bias}}{\sqrt{2}}$ & 16 & 12 \\ 
\hspace{0.5cm}
Bias resistor Johnson noise & $\sqrt{2}\cdot$ 3.1\,\pArtHz $\frac{V_\mathrm{bias}}{\sqrt{2}}$ & 15 & 11 \\
\\ 
          \em{Total} &                                                    &
         47 & 64,71,85 \\ 
                  &   &  & ($\xi=0$,$\xi=0.3$,$\xi=1$) \\
\hline 
\end{tabular} 

\caption{ \label{t_darkNoise} Theoretical noise equivalent power
  expectation for a typical SPT detector. The ``dark''
  column corresponds to the operating conditions for the noise spectra
  recorded in the laboratory and shown in Figure~\ref{f_darkNoise},
  wherein there is no incident radiation.  The ``on-sky'' column
  corresponds to typical incident radiation power for a 150~GHz SPT
  detector. Totals are included for several cases of photon correlations
  ($\xi$=0, 0.3, and 1). A similar table appears in
  Ref.~\cite{apexSZ_instrument} for APEX-SZ detectors.  
  Device and experiment related parameters are defined at the top of
  the table, $h$ is Planck's constant, $k_\mathrm{B}$ is the
  Boltzmann constant, $\xi$ is the photon correlation parameter, 
  and $\gamma_\mathrm{NE}$ is the non-equilibrium 
  parameter~\cite{Mather:82}. 
}
\end{table*}

TES detectors exhibit thermal carrier (or phonon) noise, which arises 
from random variations in the flow of phonons from the TES to the 
thermal bath. A factor $\gamma_\mathrm{NE}$ has been
included in the Table~\ref{t_darkNoise} equation describing this noise
to account for a temperature gradient along the thermal
link~\cite{Mather:82}.  We have taken 
$\gamma_\mathrm{NE}=\valGammaNE$~\cite{2011ShirokoffPhD},
consistent with a normal metal link at our operating temperatures. For
the dark detector power spectrum density shown in
Figure~\ref{f_darkNoise}, thermal carrier noise dominates above 1~Hz
and is eventually cut-off by the TES time constant $\tauTES \sim 1$\,ms, which is
not visible in the plot as it is beyond the low-pass filter cut-off.

Both the detector and bias resistor produce Johnson noise in the
narrow $LC\Rbolo$ pass-band that surrounds the carrier. The TES
detector Johnson noise current $\sqrt{4 k_\mathrm{B} T_\mathrm{bolo}/R_\mathrm{bolo}}$ is
superimposed on the amplitude modulated carrier. For a
sinusoidal-biased TES, 
this Johnson noise beats with the carrier and is mixed down to
base-band where its thermal power lies within the detector time
constant and is suppressed by electro-thermal feedback. This
suppression is similar to that which takes place for a static-biased
TES~\cite{irwin_tes:2005}, and is derived in Appendix~B of
Ref.~\cite{2011LuekerPhD}. For the operating parameters of SPT and
APEX-SZ detectors, the TES Johnson noise suppression needs to be
calculated numerically, but can be approximated by $\sim
1/\loopgain$ for the component that is in-phase with the bias 
carrier~\cite{2011LuekerPhD} (the out-of-phase component receives no
suppression, but is discarded by the demodulator). The
bias resistor suppression is less than 20\%~\cite{2011LuekerPhD} and
has been neglected in the noise expectation tabulated in
Table~\ref{t_darkNoise}. For a sinusoidal-biased detector, 
Johnson noise is enhanced by a factor $\sqrt{2}$ because uncorrelated
noise from both sidebands appear post-demodulation,
as discussed in Appendix~\ref{s_demodulator_response}. This noise is referred
to a noise equivalent power by the inverted responsivity $\sim
-V^\mathrm{RMS}_\mathrm{bias}/\sqrt{2}$. The result is that the total Johnson noise
is $\sqrt{4 k_\mathrm{B} T_\mathrm{bolo} \Pelec}/\loopgain$,
exactly as it would be for a detector biased with a static voltage.


The readout electronics contribute noise from the bias resistor,
\squid, and room temperature electronics as described below.  As
described in \S~\ref{s_squid_and_electronics}, each
multiplexer module bias resistor is located at 4~K adjacent to the
\squid s (see Figure~\ref{f_fmux_cartoon}) rather than on the sub-kelvin
stage adjacent to the detectors. This higher temperature results in
higher Johnson noise from the bias resistor. The advantage is that the number of wires
connecting to the sub-kelvin stage is just two per multiplexer comb,
rather than the four wires that would be required if the bias
resistors were located on the sub-kelvin stage. This noise
contribution could be made negligible by using a capacitive voltage
divider~\cite{2009AIPC.1185..245V}, an inductive bias, or moving the
bias resistor to lower temperature. 

The \squid s contribute a white noise level of 3.5\,pA/\rtHz.
Since the detectors are biased at high frequency, the bolometer
signals are modulated at frequencies well above the 1/f noise of the
\squid s and pre-mixer amplifiers in the room temperature system. The
room temperature electronics noise is 3.7\,pA/\rtHz, after being
referred back to the bolometer current by the \squid\ transimpedance,
$Z^\mathrm{SQ}_\mathrm{trans}$. A higher
$Z^\mathrm{SQ}_\mathrm{trans}$ would reduce this noise that is
dominated by the 1\,nV/\rtHz\ input noise of the first stage amplifier
that follows the \squid\ inside the flux locked loop. The room
temperature 
electronics noise also receives contributions from Johnson noise in
the resistors that set the gain of the amplifiers and convert voltages
to currents.

When we account for these noise contributions, the total noise
expectation is in good agreement with measurements of dark detectors
as shown in Figure~\ref{f_darkNoise}.

\subsubsection{On-Sky Noise}
\label{s_on_sky_noise}

The noise expectation for the detector shown in
Figure~\ref{f_darkNoise}, extrapolated to typical SPT 150\,GHz
observing conditions with 11\,pW of incident radiation power, is shown
in the ``on-sky'' column of Table~\ref{t_darkNoise}. This radiation
power results in a reduction of the Joule heating necessary to keep
the detector at 0.73\,$R_\mathrm{Normal}$, so that the detector is
tuned with a lower bias voltage. 
\revisedtext{ This lower bias voltage results in a reduced NEP for all 
  current noise sources, such as the three components of readout 
  noise listed in Table~\ref{t_darkNoise}. }
Since the low frequency noise from the readout system is proportional
to the bias voltage amplitude, its amplitude is also reduced. If
readout were the only contribution to low-frequency noise, the knee
would be at 0.4~Hz for this detector parameters.

In addition to the thermal carrier noise, ETF suppressed Johnson
noise, and readout noise, fluctuations in the rate of incident photons
produces an additional noise term corresponding to photon shot noise
and photon
correlations~\cite{1986ApOpt..25..870L,2003ApOpt..42.4989Z}, also
included in Table~\ref{t_darkNoise} for the ``on-sky'' detector. The
degree of photon correlation depends on poorly characterized details
of the filters and coupling,
so that the coefficient $\xi$~\cite{2003PhDT........16R} is often
introduced to parametrize a correction factor for the simple equation
presented in Table~\ref{t_darkNoise}. $\xi$ lies between 0-1, with 0.3
being a reasonable choice. Table~\ref{t_darkNoise} includes noise
expectations spanning the full range.

\begin{figure}
\includegraphics[width=0.475\textwidth,clip=]{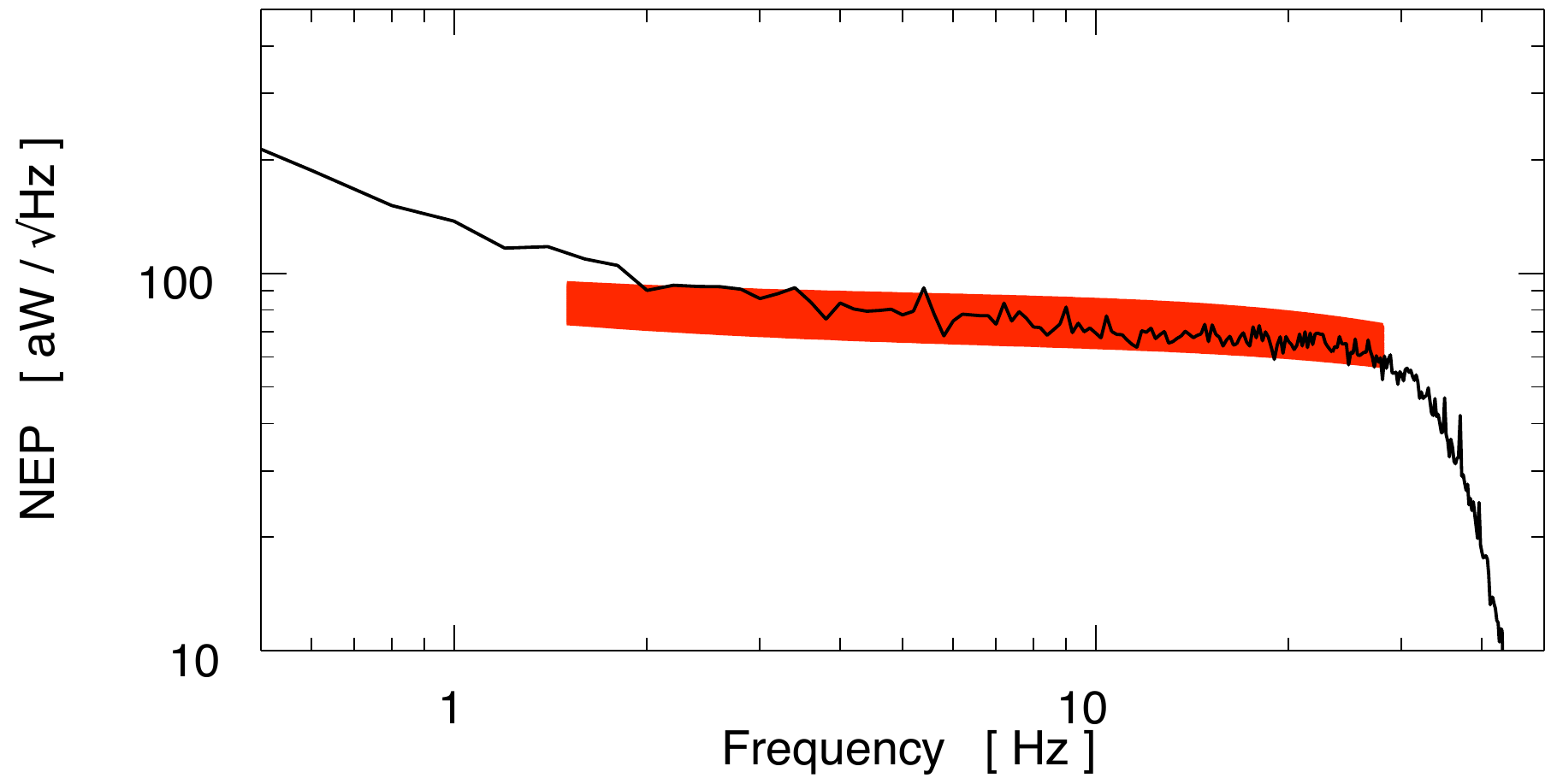}
\caption{ \label{f_lightDetectorNoise} 
Power spectrum distribution of a bolometer 
observing in the 150~GHz band on the South Pole Telescope is shown. 
The low frequency noise is primarily due to atmospheric fluctuations. 
This detector's 
performance is typical for detectors in the 150\,GHz band and agrees 
well with the expectation, shown as a shaded band. Its measured 
noise equivalent temperature (NET$_\mathrm{CMB}$) is 409\,$\mu \mbox{K} \cdot
\sqrt{\mbox{s}}$ projected on the sky. 
}
\end{figure}

The power spectrum distribution for a representative 150~GHz bolometer 
operating ``on-sky'' during the 2009 observing season on the South
Pole Telescope is shown in Figure~\ref{f_lightDetectorNoise}. This
data corresponds to the raw timestream; no filtering has been applied
to remove atmospheric signals or other effects. The SPT bolometer
stage was operated at a bath temperature of 
$T_\mathrm{bath}=$260\,mK and the detector
 has sufficiently similar device parameters to the dark
detector, shown in Figure~\ref{f_darkNoise}, that its total noise
expectation is the same to within 2\% of the values presented in
Table~\ref{f_darkNoise}. Superimposed on
Figure~\ref{f_lightDetectorNoise} is a shaded band corresponding to the
noise expectation for this detector, excluding atmospheric contributions. The width of the shaded band
indicates the range of allowed photon correlations $0<\xi <1$.

At frequencies below $\sim$3\,Hz, the spectrum is dominated by
atmospheric fluctuations that are not included in the theoretical
expectation presented in Table~\ref{t_darkNoise}. For science
analyses, filtering and subtraction techniques (see e.g.,
\cite{2008arXiv0810.1578S}) applied in the data processing pipeline
are used to reduce the effect of atmospheric signals to below 1~Hz.
Above this frequency, the spectrum agrees well with the expectation.
After calibrating this detector response with astrophysical sources,
its measured sensitivity is 409\,$\mu \mbox{K}_\mathrm{CMB} \cdot
\sqrt{\mbox{s}}$ projected on the sky and referred to a source at the
CMB temperature. This is typical of SPT 150\,GHz detector performance.
Noise and sensitivity statistics for the APEX-SZ and SPT receiver
systems are presented in Refs.\,\cite{BensonSptReceiver} and
\cite{apexSZ_instrument} respectively.



\section{Conclusions} \label{s_conclusions}

We have presented the design and performance of the SQUID-based frequency-domain
multiplexed (\fMUX) readout system that has been deployed for TES detectors on
the APEX-SZ instrument and the South Pole Telescope SZ-camera. The readout reduces the heat-load on the
sub-kelvin cryogenic stage by sharing wiring among many detectors
using a simple architecture that minimizes the sub-kelvin system complexity.
It has near-zero power dissipation on the sub-kelvin stage, is extensible
to much higher multiplexing factors, allows bias levels to be set
individually for each detector, is modular with no shared
components between readout modules, and exhibits strong reduction of
microphonic response and low-frequency EMI because the sky signals are
modulated up to high frequency.

The system has been characterized in terms of noise performance,
sensitivity, detector yield, and cross-talk in both the laboratory and
with on-sky observations. The measured noise equivalent power of the system
is consistent with theoretical expectations, and contributions from the
readout system are small in comparison to detector and photon noise
above 1~Hz. At low frequencies, the system noise is dominated by
residual atmospheric fluctuations. 

These implementations have shown that frequency-domain multiplexing is
an effective technology for scaling to very large arrays of TES
detectors. An upgraded digital version of the analog \fMUX\ back-end
electronics has recently been developed~\cite{2008ITNS...55...21D} and
has been deployed for the EBEX~\cite{2010SPIE.7741E..52A}
balloon-borne CMB polarimeter and the ground-based
POLARBEAR~\cite{2010SPIE.7741E..39A} CMB polarimeter. \mynewtext{ A
  polarimeter camera~\cite{2009AIPC.1185..511M} for the South Pole
  Telescope is being commissioned during the austral 2011/2012 summer
  and uses the digital \fMUX. } The digital \fMUX\ system achieves an
order of magnitude lower power consumption, higher multiplexing
factors, faster detector and \squid\ setup, better low frequency noise
performance, and lower cost. Development is underway to develop \fMUX\
electronics for satellite applications~\cite{2011_FMUX_FOR_FLIGHT}.
\revisedtext{
Present prototypes use Digital Active
Nulling~\cite{2012_DAN_DAF} in place of the flux-locked loop. This
extends the \squid\ bandwidth to allow for 64 bolometers per \squid\
module using a single pair of sub-Kelvin wires.
}

\begin{acknowledgments}
  We thank Andy Smith for useful discussions and design/fabrication of
  niobium inductors at Northrup Grumman, Kent Irwin and Gene Hilton
  for useful discussions and design/fabrication of the series array
  \squid s at NIST, and Hannes Hubmayr, Bryan Steinbach, Kam Arnold,
  and Graeme Smecher for useful discussions of bolometer responsivity
  and demodulator response.

  The National Science Foundation funds APEX-SZ through grants
  AST-0138348 \& AST-0709497 and the South Pole Telescope through
  grants ANT-0638937 and ANT-0130612. Work at LBNL is supported by
  the U.S.\ Department of Energy under Contract No. DE-AC02-05CH11231.
  The McGill team acknowledges funding from the Natural Sciences and
  Engineering Research Council of Canada, Canadian Institute for
  Advanced Research, and the Canadian Foundation for Innovation. MD
  acknowledges support from the Canada Research Chairs program and a
  Sloan Fellowship. 
\end{acknowledgments}

\appendix
\section{Modulation and demodulation transfer functions for alternating-voltage biased bolometers}
\label{s_demodulator_response}

In this section, we consider the transfer functions for the modulation
and demodulation of signals for a bolometer voltage biased with a
constant amplitude sinusoidal carrier. We also consider the transfer
function for noise that is superimposed on the amplitude-modulated
carrier.

First consider a single Fourier component of a small power signal $P_s
\cos \omega_s t$ absorbed by a bolometer that is biased deep in its
transition with a constant amplitude voltage bias $V_\mathrm{bias}=V_c \sin
\omega_c t$.

Under the assumption of strong electro-thermal feedback (ETF), the
total power on the detector will be constant and determined by the
average thermal conductivity $\bar{G}$ and temperature difference from
the bolometer to its bath $\Delta T= T_\mathrm{bolo} - T_\mathrm{bath}$,
\begin{equation} \label{e_power_balance} 
P_\mathrm{TOTAL} = \bar{G} \Delta T
              = V_\mathrm{bias} I_\mathrm{bolo} + \Prad + P_s \cos \omega_s t ,
\end{equation}
where $\Prad$ is the approximately constant radiation loading
power from the sky.
The amplitude of the bolometer voltage bias is constant, so that the power
signal produces an amplitude modulation $I_s$ on the bolometer carrier
current $I_c$:
\begin{equation}
  I_\mathrm{bolo} = (I_c + I_s \cos \omega_s t) \sin \omega_c t .
\end{equation}
Keeping only the time varying terms from Equation~\ref{e_power_balance},
the power balance is
\begin{equation}
  - P_s \cos \omega_s t
              = V_c I_s \cos \omega_s t \sin^2 (\omega_c t) .
\end{equation}
Since the carrier frequency $\omega_c$ is much faster than the bolometer
time constant, the $\sin^2 (\omega_c t)$ factor integrates to 1/2,
leading to
\begin{equation}
  - P_s = \frac{V_c}{2} I_s = \frac{V^\mathrm{RMS}_c}{\sqrt{2}} I_s .
\end{equation}
Thus, the strong ETF responsivity for a bolometer biased with a
sinusoidal voltage is
\begin{equation} \label{e_ave_responsivity_infiniteL}
  \frac{d I_s}{dP_s} = -\frac {\sqrt{2}}{V^\mathrm{RMS}_c}
\end{equation}
for amplitude modulated signals. 
For simplicity, we have left out the factor that accounts for the bolometer 
time constant.
Equation~\ref{e_ave_responsivity_infiniteL} should be compared to 
$\frac{d I_s}{dP_s} = -\frac {1}{V_c}$ for 
a constant-voltage biased bolometer~\cite{Lee:98}. For finite loop-gain 
\loopgain, the responsivity in
Equation~\ref{e_ave_responsivity_infiniteL} becomes 
\begin{equation} \label{e_av_responsivity}
  \frac{d I_s}{dP_s} = -\frac {\sqrt{2}}{V^\mathrm{RMS}_c}
  \frac{\loopgain}{1+\loopgain} .
\end{equation}

Next we consider a bias carrier waveform $I_c \sin \omega_c
t$ and present the response functions for sine-wave demodulation (the
traditional mixer reference waveform) and square-wave demodulation
(the mixer reference used for this implementation of the \fMUX\
system) as a function of the phase difference between the carrier and
the demodulator, $\phi_\mathrm{ref}$. The results for in-phase $I$
demodulation and 90-degree out-of-phase $Q$ demodulation are
summarized in Table~\ref{t_mixerResponse}.

\begin{figure}
\includegraphics[width=0.45\textwidth,clip=]{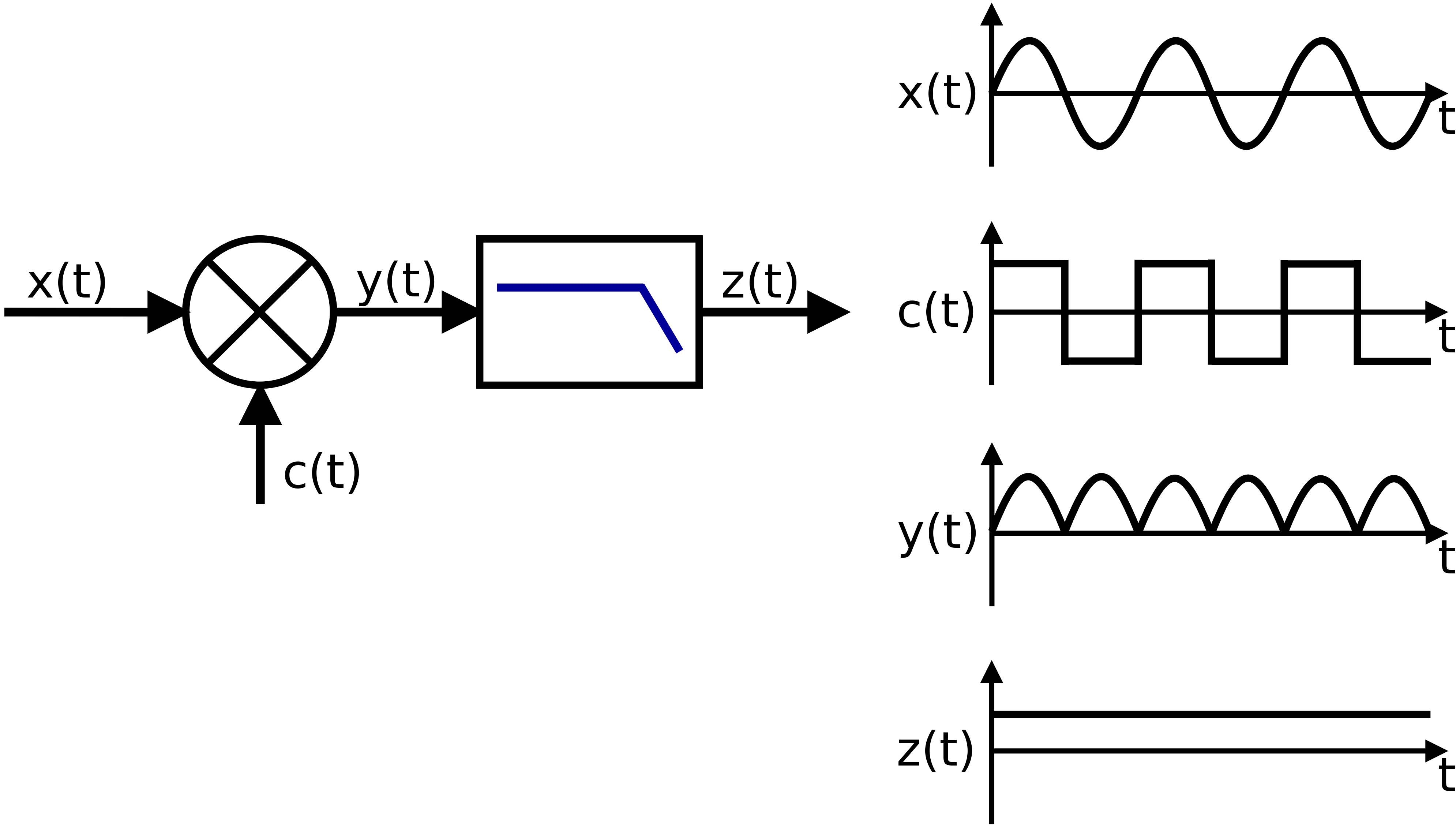}
\caption{ \label {f_genericMixer} A generic demodulator, consisting of
  a mixer and low-pass filter is shown (left). Waveforms for the
  special case of a sinusoidal input demodulated with a square wave
  reference is shown at right.}
\end{figure}

A generic demodulator is shown in Figure~\ref{f_genericMixer} (left),
where an input signal $x(t)$ is multiplied by a reference signal
$c(t)$ then integrated with a low-pass filter (LPF) to produce the
output $z(t)$. The cutoff frequency of the LPF is much lower than
the reference frequency and much larger than the modulation-encoded
sky-signals.

 For the special case of an on-resonance, in-phase
sinusoidal input $x(t)=I_c \sin(\omega_c t)$ and a unit amplitude
sine-wave reference $c(t)=\sin(\omega_c t)$, the mixer output is the
familiar $y(t)=x(t)\times c(t) = I_c \sin^2(\omega_c t) = I_c \frac{1 - \cos
  2\omega_c t}{2}$. The LPF averages the second term to zero, so that
the demodulator baseband output is $z(t)=I_c /2$.

The $x(t),~c(t),~y(t)$, and $z(t)$ waveforms are shown for this
special case of a square-wave mixer in Figure~\ref{f_genericMixer}
(right). The reference is an on-resonance, in-phase, unit amplitude
square wave $c(t)$. The output is a baseband signal with amplitude equal to
the average of a rectified sine wave, $\frac{2}{\pi} I_c$. If the
phase of the reference differs from the sinusoid phase by
$\phi_\mathrm{ref}$, the amplitude of the baseband output signal
$z(t)=\frac{2}{\pi} I_c \cos \phi_\mathrm{ref}$, meaning that the response
function is $\frac{2}{\pi}$ for in-phase $I$ demodulation and zero for
$Q$.

\begin{figure}
\includegraphics[width=0.45\textwidth,clip=]{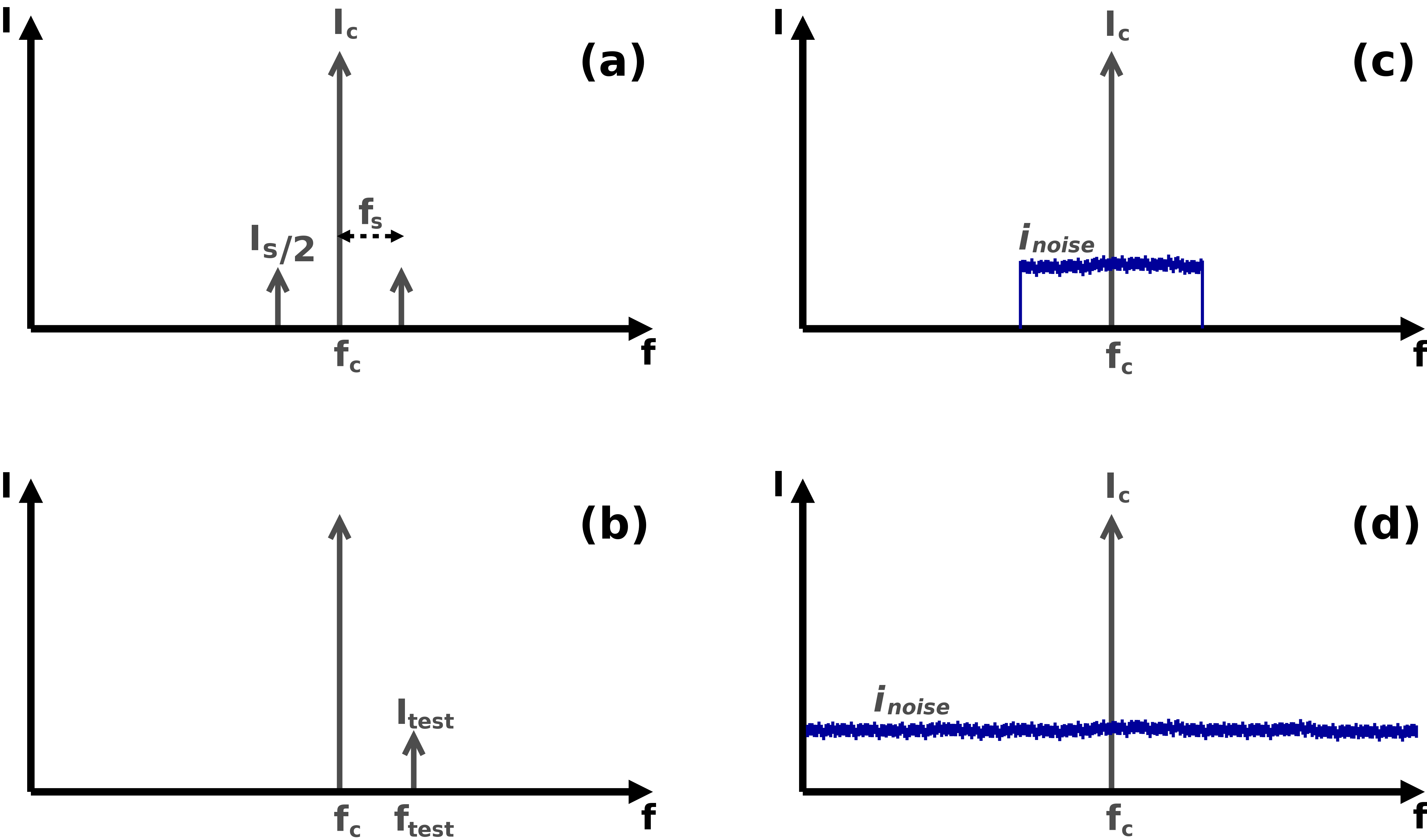}
\caption{ \label{f_mixerResponse} Frequency-domain representations of
  (a) a modulation of carrier amplitude $I_c$ with a signal of
  amplitude $I_s$, (b) an uncorrelated sinusoid with amplitude
  $I_\mathrm{test}$ adjacent to carrier amplitude $I_c$, (c) narrow-band white
  noise with amplitude $i_{\mbox{noise}}$ superimposed at frequencies
  surrounding carrier amplitude $I_c$, and (d) broadband white noise
  with amplitude $i_{\mbox{noise}}$ superimposed on carrier amplitude
  $I_c$. A fifth case, where narrow-band white noise modulates the
  carrier (as would be the case for bolometer phonon or photon noise)
  is not shown.}
\end{figure}

Figure~\ref{f_mixerResponse} shows other demodulator input signals
that are relevant for understanding the signal and noise transfer
functions of the \fMUX\ system. The first case (a) corresponds to a bias carrier 
$I_c \sin \omega_c t,\ \omega_c \gg \omega_\mathrm{LPF}$ 
modulated by a small sky signal 
$I_s \cos \omega_s t,\ \omega_s \ll \omega_\mathrm{LPF}$, 
yielding a demodulator input signal
\begin{eqnarray}
x(t) =   && (I_c + I_s \cos \omega_s t) \sin \omega_c t \\
      = && I_c \sin \omega_c t 
           + \frac{I_s}{2} \left[ \sin (\omega_c+\omega_s) 
                                                  + \sin (\omega_c-\omega_s) \right]. \nonumber
\end{eqnarray}
This corresponds to the amplitude modulation produced by a power
signal discussed in the first part of this Appendix; and arises from
sky signals, bolometer phonon noise, and bolometer photon noise.
The sky signal is split into correlated sidebands adjacent to the
carrier and appears in the demodulated timestream with amplitude
$z(t)=\frac{2}{\pi} I_s \cos \phi_\mathrm{ref} \cos \omega_s t$.

Case (b) in Figure~\ref{f_mixerResponse} corresponds to an off-resonance sinusoid 
$I_t \sin \omega_t t$ injected adjacent to the carrier. For
simplicity, this test sinusoid is assumed to be injected after the
bolometer so that the sinusoid does not deposit power in the
bolometer that would be subject to ETF. The test sinusoid beats with
the carrier 
so that the demodulated signal is 
$z(t)=\frac{1}{2} I_t \cos \phi_\mathrm{ref} \sin | \omega_t - \omega_c | t$
for a sine-wave mixer and
$z(t)=\frac{2}{\pi} I_t \cos \phi_\mathrm{ref} \sin | \omega_t - \omega_c | t$ 
for a square-wave mixer.
Notice that the test signal has the same demodulation attenuation factor as an
on-resonance amplitude modulated signal. Test signals of this type are
used extensively in the \fMUX\ system for measuring electronics
transfer functions and characterising detectors.

The situation for white noise added to the carrier waveform
 (i.e., does not amplitude-modulate the carrier) requires special
 attention, because it differs from noise that amplitude modulates the
 carrier and because the square-wave mixer responds differently to
 broadband and narrow-band noise sources.

 Case (c) in Figure~\ref{f_mixerResponse} corresponds to bandwidth
 limited white noise adjacent to the carrier with amplitude $x(t) = i_{noise}
 [$A/\rtHz $]$. An example would be Johnson noise from a resistor in
 an $LCR$ notch filter (or from a normal bolometer in our \fMUX\
 circuit, though ETF complicates the situation for a bolometer in
 transition, suppressing this noise). A single sideband can be
 decomposed into sinusoids, and hence has the same response function
 as case (b). There are, however, contributions from the sidebands on
 either side of the carrier, each contributing the same power. The two
 sidebands are uncorrelated with one another, resulting in a $\sqrt{2}$
 enhancement, for a total contribution $z(t)=\frac{\sqrt{2}}{2}
 i_\mathrm{noise}$ for a sine-wave mixer and 
$z(t)=\frac{2\sqrt{2}}{\pi}
 i_\mathrm{noise}$ for a square-wave mixer. There is no dependence on
 $\phi_\mathrm{ref}$ since this input signal is not coherent. The end
 result is that the sky-signal to narrow-band noise ratio is the same
 for a TES read out through a square-wave mixer and a sine-wave
 mixer, and equivalent to what would be obtained for a TES with a static
 bias.

 Case (d) in Figure~\ref{f_mixerResponse} corresponds to broadband
 white noise with amplitude $i_\mathrm{noise} [$A/\rtHz $]$. An example would
 be \squid\ noise or readout electronics noise. For a sine-wave mixer,
 the response is identical to that of case (c). For a square-wave
 mixer, both the noise power adjacent to the carrier {\it and} the
 noise power adjacent to the odd-harmonics of the carrier contribute.
 This is immediately evident by examining the frequency-space
 representation of the square-wave, with includes delta functions at
 all of the odd-harmonics that fall in amplitude as $1/n$.  Since the
 noise waveform is characterized by random phase and the square-wave
 mixer simply rotates the phase of half the time samples by 180\degree,
 it is easy to see that the demodulated amplitude will be unchanged,
 $z(t)=i_\mathrm{noise}$. We note that the demodulated amplitude of
 broadband noise can be reduced to that of case (c) if the carriers
 span less than 2 octaves by placing a low-pass filter in front of the
 mixer, removing input from the 3$^{rd}$ harmonics.
 The end
 result is that the sky-signal to narrow-band noise ratio is the same
 for a a TES readout out through a square-wave mixer and a sine-wave
 mixer, and equivalent to what would be obtained for a TES with a static
 bias. The end
 result is that the sky-signal to broadband noise ratio is 
 $(2/\pi)/(1/\sqrt{2})\simeq 0.90$ worse for a TES read out
  through a square-wave mixer as compared to a sine-wave mixer. The
  latter is equivalent to what would be obtained for a TES with a static
  bias. Typically for \fMUX\ readout, the contribution from broadband
  noise sources is small, and this difference is negligible.

\begin{widetext}
\begin{table*}\renewcommand{\arraystretch}{1.5}\addtolength{\tabcolsep}{2pt}
\begin{tabular}{lc|cc|cc}
               & & \multicolumn{2}{c|}{$c(t)= $Square-wave reference} 
                   & \multicolumn{2}{c}{$c(t)=\sin(\omega_c t)$ reference} \\
Input Signal $x(t)$ & & $I$ & $Q$ 
                                 & $I$ & $Q$ \\ \hline
Sinusoidal carrier & $I_c \sin \omega_c t$ 
                            & $\frac{2}{\pi} I_c$ & 0 
                            & $\frac{1}{2} I_c$    & 0       \\
Amplitude modulated carrier $(a)$ & $ (I_c + I_s \cos \omega_s t) \sin \omega_c t$
                            & $\frac{2(I_c + I_s \cos \omega_s t)}{\pi}$ & 0
                            & $\frac{I_c + I_s \cos \omega_s t}{2}$ & 0 \\
Sinusoidal test signal $(b)$ & $I_t \sin \omega_t t,\ \omega_t\ne\omega_d$
     & \multicolumn{2}{c|}{$\dagger \frac{2}{\pi} I_t \sin (\omega_t-\omega_c) t$}                  
     & \multicolumn{2}{c}{$\dagger \frac{1}{2} I_t \sin (\omega_t-\omega_c) t$} \\
Superimposed narrowband white noise $(c)$ & $n(t)  $
     & \multicolumn{2}{c|}{$\frac{2\sqrt{2}}{\pi} n(t)  $}                  
     & \multicolumn{2}{c}{$\frac{\sqrt{2}}{2} n(t)  $} \\
Superimposed broadband white noise $(d)$ & $n(t)  $
     & \multicolumn{2}{c|}{$n(t)  $}                  
     & \multicolumn{2}{c}{$\frac{\sqrt{2}}{2} n(t)$} \\

\hline
\end{tabular}
\caption{\label{t_mixerResponse} 
  Demodulator response factors are tabulated for various signal and noise types including those shown in panels (a) through (d) of Figure~\ref{f_mixerResponse}. $\omega_c$ is the demodulator reference frequency. Notice that white noise that is superimposed on the carrier (not amplitude modulated) has a factor $\sqrt{2}$ enhancement over amplitude modulated signals. For the case of a square mixer and broadband white noise, the factor is $\frac{\pi}{2}$ because noise located within the LPF bandwidth of the reference signal's odd harmonics also contribute.
  \\ $\dagger$ An arbitrary phase which depends on the start time is omitted from these equations.
}
\end{table*}
\end{widetext}




\end{document}